\documentclass[sigconf]{acmart}

\AtBeginDocument{%
  }
    
\copyrightyear{2026}
\acmYear{2026}
\setcopyright{cc}
\setcctype{by}
\acmConference[CHI '26]{Proceedings of the 2026 CHI Conference on Human Factors in Computing Systems}{April 13--17, 2026}{Barcelona, Spain}
\acmBooktitle{Proceedings of the 2026 CHI Conference on Human Factors in Computing Systems (CHI '26), April 13--17, 2026, Barcelona, Spain}
\acmPrice{}
\acmDOI{10.1145/3772318.3790815}
\acmISBN{979-8-4007-2278-3/2026/04}

\usepackage{amsmath}
\usepackage{algorithm}
\usepackage{makecell}

\usepackage{color}
\usepackage{enumitem}
\usepackage{bbding}
\usepackage{booktabs}
\usepackage{graphicx}
\usepackage{amsmath, bm}
\usepackage{enumitem}
\setlist{noitemsep,parsep=0pt,partopsep=0pt, leftmargin=10pt} 
\usepackage{listings}
\usepackage{multirow}
\usepackage{colortbl}

\usepackage{tabu}                      
\usepackage{booktabs}                  
\usepackage{lipsum}                    
\usepackage{mwe}                       
\usepackage{tabularx}
\usepackage{ragged2e}
\usepackage{array}
\usepackage{multirow}

\definecolor{mygreen}{RGB}{0,176,80}
\definecolor{cback}{HTML}{EDEFF1}
\definecolor{cframe}{HTML}{B9C4CA}
\definecolor{cgrey}{HTML}{666666}
\newcommand{\tool}{\textit{DiLLS}\xspace}
\newcommand{\ie}{i.e.{\xspace}}
\newcommand{\eg}{e.g.,\xspace}

\definecolor{tablerowcolor}{rgb}{0.667,0.667,0.667 }
\definecolor{tablerowcolor2}{rgb}{0,0,0}
\definecolor{visual}{HTML}{e8efd9}
\definecolor{motion}{HTML}{fde7d5}
\definecolor{narrative}{HTML}{e2dce9}
\definecolor{audio}{HTML}{d6ebf2}
\definecolor{bluecrayola}{rgb}{0.12,0.46,1.0}
\definecolor{Blueberry}{RGB}{79,104,196}

\usepackage{hyperref}

\definecolor{highlight}{HTML}{1155cc}
\definecolor{review}{HTML}{f72585}
\definecolor{highlight}{HTML}{000000}
\usepackage{marginnote}
\global\marginparsep=8pt
\definecolor{highlight}{HTML}{0077b6}
\definecolor{highlight}{HTML}{000000}
\newcommand{\rui}[1]{\textcolor{highlight}{#1}}
 
\begin{document}

\title{\tool: Interactive Diagnosis of LLM-based Multi-agent Systems via Layered Summary of Agent Behaviors}

\author{Rui Sheng}
\authornote{Equal Contribution}
\orcid{0000-0001-9321-6756}
\affiliation{%
  \institution{The Hong Kong University of Science and Technology}
  \city{Hong Kong}
  \country{China}
}
\email{rshengac@connect.ust.hk}

\author{Yukun Yang}
\authornotemark[1]
\orcid{0009-0003-1971-4468}
\affiliation{%
  \institution{Tongji University}
  \city{Shanghai}
  \country{China}
  }
\email{yangyyk@tongji.edu.cn}

\author{Chuhan Shi}
\orcid{0000-0002-3370-1626}
\affiliation{%
  \institution{Southeast University}
  \city{Nanjing}
  \country{China}
}
\email{cshiag@connect.ust.hk}

\author{Yanna Lin}
\orcid{0000-0003-3730-0827}
\affiliation{%
  \institution{University of Waterloo}
  \city{Waterloo}
  \country{Canada}
}
\email{yanna.lin@uwaterloo.ca}

\author{Zixin Chen}
\orcid{0000-0001-8507-4399}
\affiliation{%
  \institution{The Hong Kong University of Science and Technology}
  \city{Hong Kong}
  \country{China}
}
\email{zchendf@connect.ust.hk}

\author{Huamin Qu}
\orcid{0000-0002-3344-9694}
\affiliation{%
  \institution{The Hong Kong University of Science and Technology}
  \city{Hong Kong}
  \country{China}
}
\email{huamin@cse.ust.hk}

\author{Furui Cheng}
\orcid{0000-0003-2329-6126}
\authornote{Corresponding Author}
\affiliation{%
  \institution{ETH Zürich}
  \city{Zürich}
  \country{Switzerland}}
\email{furui.cheng@inf.ethz.ch}
\renewcommand{\shortauthors}{Trovato et al.}

\begin{abstract}
Large language model (LLM)-based multi-agent systems have demonstrated impressive capabilities in handling complex tasks. However, the complexity of agentic behaviors makes these systems difficult to understand. When failures occur, developers often struggle to identify root causes and to determine actionable paths for improvement.
Traditional methods that rely on inspecting raw log records are inefficient, given both the large volume and complexity of data. 
To address this challenge, we propose a framework and an interactive system, \tool{}, designed to reveal and structure the behaviors of multi-agent systems.
The key idea is to organize information across three levels of query completion: activities, actions, and operations. By probing the multi-agent system through natural language, \tool{} derives and organizes information about planning and execution into a structured, multi-layered summary.
Through a user study, we show that \tool{} significantly improves developers’ effectiveness and efficiency in identifying, diagnosing, and understanding failures in LLM-based multi-agent systems.

\end{abstract}


\begin{teaserfigure}
  \includegraphics[width=\textwidth]{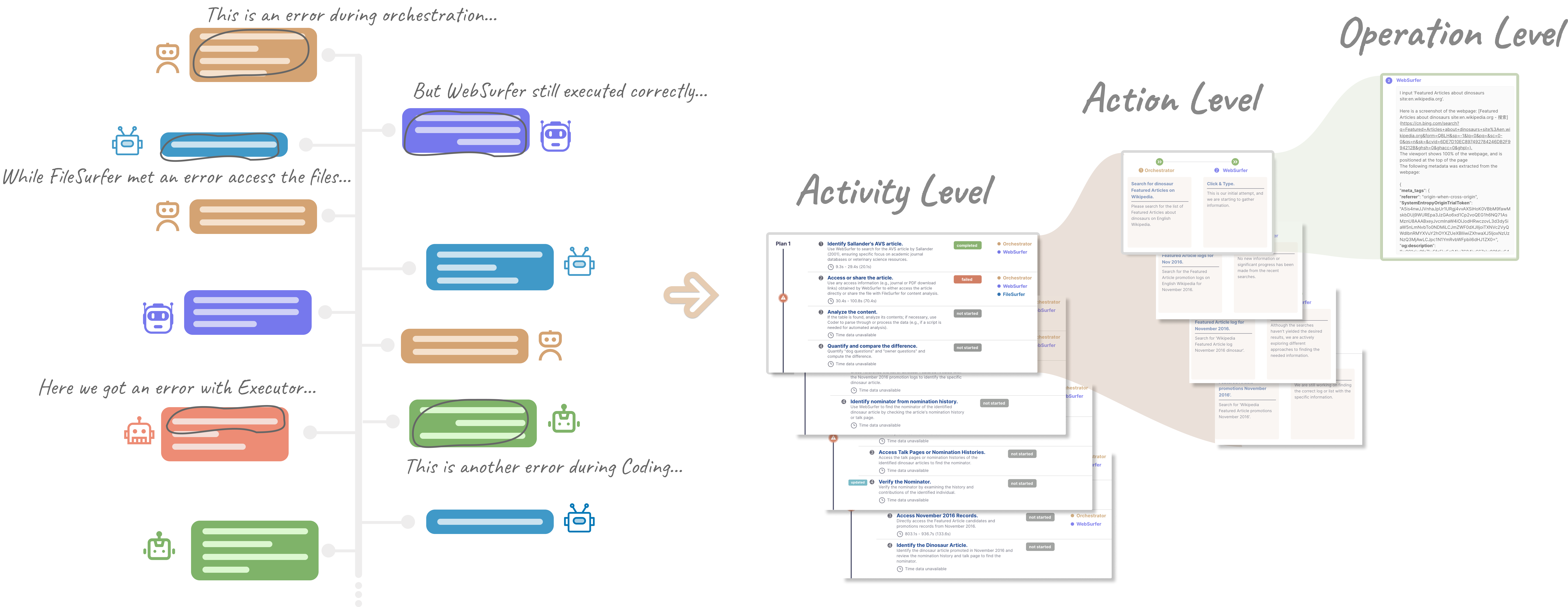}
  \caption{We designed a layered exploration approach and implemented a system, \tool, to assist AI developers in exploring agent behaviors across various layers for multi-agent system diagnosis.}
  \Description{A conceptual figure illustrating the three-layered approach of our system for diagnosing multi-agent systems. The figure shows three layers: ``Activity Level'' (top), ``Action Level'' (middle), and ``Operation Level'' (bottom). Additionally, a screenshot of the system interface is shown, with callouts linking specific interface elements to these three conceptual layers, demonstrating how the system translates low-level data into high-level insights.}
  \label{fig:teaser}
\end{teaserfigure}


\maketitle
\section{Introduction}

There is a growing interest in organizing large language models (LLMs) into agentic architectures to handle complex tasks~\cite{wang2024survey, Lin2025survey}. 
Such multi-agent systems harness the capabilities of LLMs to interpret human instructions, plan execution strategies, and delegate responsibilities across individual agents. 
When further augmented with features such as short-term or long-term memory and external tool use, these agents can collaborate more effectively to solve complex problems. 
Promising applications have already emerged across domains including web searching~\cite{fourney2024magentic, ala2025autonomous}, scientific research~\cite{GAO2024Empowering, Ghafarollahi2024SciAgents, GHAFAROLLAHI2024ProtAgents}, and software development~\cite{manish2024autonomous, rasheed2024codepori}. 

While LLM-based multi-agent systems introduce promising capabilities in autonomous task solving, monitoring and diagnosing their behaviors remain challenging~\cite{epperson2025interactive}. 
Unlike traditional autonomous systems with human-authored running logic and well-defined execution paths, 
LLM-based agentic systems operate with a high degree of flexibility in task scheduling and execution, making their behaviors hard to track, interpret, and regularize. 
The involvement of multiple agents further amplifies this complexity, introducing additional challenges in coordinating and understanding their interactions.
When failures occur, it is often unclear how to identify, understand, or resolve the underlying issues. 

Conventional debugging relies on scanning agent logs and execution histories, 
but in LLM-based systems, interactions are typically conducted through natural language dialogues between agents. As a result, the logs are highly unstructured and verbose, frequently spanning hundreds of exchanges. 
Reading through such transcripts is not only time-consuming but also makes it difficult to discern agent intentions, coordination strategies, or failure points.
Additionally, the diversity of agent behaviors, ranging from web searching, coding, to file reading, can overwhelm developers with excessive details and obscure the context of the whole problem-solving process. This further complicates the task
of diagnosing and understanding system failures.
Although agent management platforms such as AutoGen Studio~\cite{dibia2024autogen}, CrewAI~\cite{crewAI}, Voiceflow~\cite{voiceflow}, LangGraph~\cite{LangchainAi}, and LangSmith~\cite{mavroudis2024langchain} provide visibility into agent execution histories, deriving root causes of failures and generating hypotheses about how to improve systems remains an open challenge.

To better understand the practical challenges of diagnosing LLM-based multi-agent systems, we conduct a formative study with eight experienced developers. 
Our findings reveal that those experts strongly demand high-level summaries of agents’ planning and execution behaviors to gain an overview of potential failure causes.
Moreover, an efficient interactive visualization interface is needed to help developers locate erroneous execution histories and interpret system failures more efficiently.

Building on insights from our formative study, we propose a framework to derive, organize, and present agentic behaviors within a centralized architecture.
The major challenge lies in determining the appropriate layers of abstraction that provide a coherent and structured organization of behaviors.
We were inspired by Activity Theory (AT), a framework for understanding human behaviors and practices, where an activity is decomposed into three levels: the overarching activity, the actions as goal-directed processes supporting the activity, and operations as the concrete executions.
Inspired by this hierarchy, we employ natural language probing techniques to analyze multi-agent system behaviors across three layers: overall planning and execution, action arrangement and outcomes, and logs for concrete operations.
Based on this layered representation, we developed a tool, \tool, to facilitate the exploration and interpretation of agent behaviors across these three levels.

We evaluate \tool through a user study with 12 multi-agent system developers. Participants were asked to diagnose an erroneous multi-agent system using both \tool and a baseline system simplified from \tool that only presents basic operation-level logs, report their findings, and provide both qualitative and quantitative feedback.
Results show that our system achieves significantly higher efficiency in error identification, as evidenced by the greater number of potential model errors detected by participants. Moreover, \tool leads to lower mental demand and workload, demonstrating its effectiveness in supporting the diagnosis of complex multi-agent behaviors. Our contributions include:

\begin{itemize}
    \item We conduct a formative study with eight AI developers to uncover how practitioners currently diagnose failures in multi-agent systems and the challenges they face.
    \item We propose a framework and a system, \tool, that structures behavioral information in multi-agent systems to support developers in understanding and diagnosing case-level failures.
    \item We validate the effectiveness of \tool through a user study, showing that it enables practitioners to identify failures more confidently and comprehensively while reducing their stress.
\end{itemize}
\section{Background and Related Work}
In this section, we first introduce the background of LLM-based multi-agent systems, including their working mechanisms and design components. Specifically, we focus on the four types of interaction architectures among agents. After that, we elaborate on the current diagnostic tools for agentic systems, clarifying their limitations as well as the innovations and scope of our tool.

\subsection{LLM-Based Multi-Agent Systems}
Advances in large language models (LLMs) have catalyzed the development of multi-agent systems capable of tackling complex tasks across diverse domains, such as web search~\cite{nie2024hybrid}, scientific research~\cite{talebirad2023multi}, software development~\cite{he2024llm,fan2024contextcam,jin2024teach}, and societal and classroom simulations~\cite{wan2024building,park2023generative,liu2024classmeta,shaikh2024rehearsal,hwang2024whose}. 
These systems typically comprise multiple specialized agents, each with distinct roles, prompts, or tools. 
They collaborate to decompose tasks, delegate subtasks, and leverage external resources or memory~\cite{park2023generative,li2024survey,zhang2024see,huang2023memory}. 
For example, Magentic-One, a multi-agent system developed by Microsoft, features an Orchestrator that directs specialized agents like WebSurfer, FileSurfer, and Coder to accomplish intricate tasks~\cite{fourney2024magentic}. A practical application could be automating research: the WebSurfer gathers web-based information, the Coder analyzes the data, and the FileSurfer compiles the findings into a structured report.
By distributing workloads and combining complementary capabilities, multi-agent approaches often outperform single-agent counterparts in addressing complex use queries~\cite{chen2023agentverse,li2025new,zhang2024see,shen2024data}.

To streamline the construction and orchestration of these LLM-based multi-agent systems, multiple frameworks have been proposed~\cite{wu2023autogen, li2023camel,wu2024copilot, crewAI, LangchainAi}. 
One of the widely accepted agent design frameworks was introduced by Li et al.~\cite{li2024survey}, which defines five key components: the creation of various agents (profile), the ability to perceive the environment (perception) and communicate with other agents (mutual interaction), the ability to respond to and complete complex tasks (self-action), and the ability to enhance their intelligence through self-reflection (evolution).

\begin{figure*}[ht]
\centering
\includegraphics[width=\linewidth]{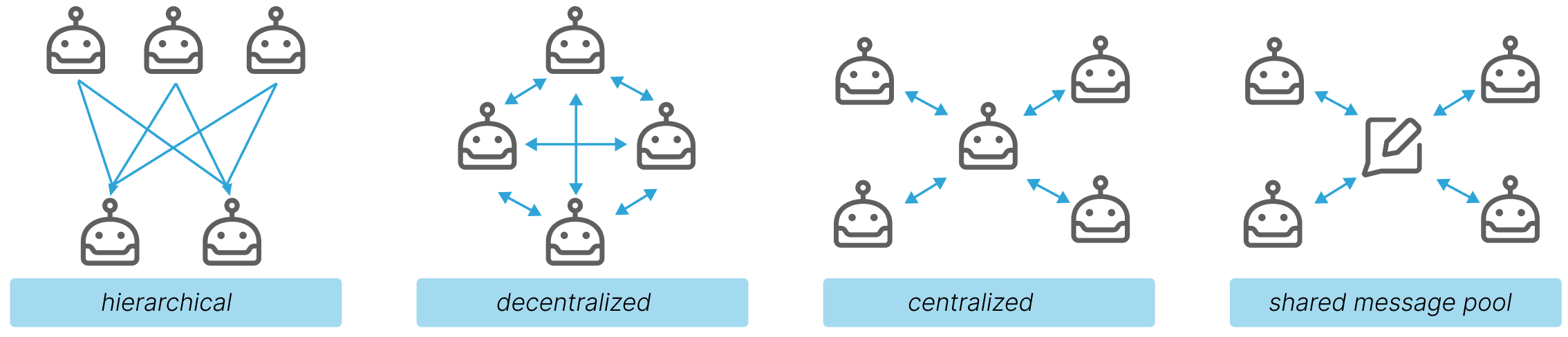}
\caption{The four interaction architectures.} 
\Description{A comparative visualization of four agent communication structures. (A) Hierarchical: agents are organized according to their responsibilities and control relationships at different levels. Tasks are typically assigned from top to bottom, and status feedback is provided from bottom to top. (B) Decentralized: a mesh network where agents communicate directly. (C) Centralized: a star topology with a central orchestrator connecting to all worker agents. (D) Shared Message Pool: agents surrounding a central database representing a common communication space.}
\label{fig: architectures}
\end{figure*}

Additionally, Li et al.~\cite{li2024survey} also summarize four types of architectures for agent interactions: hierarchical, decentralized, centralized, and shared message pool (\autoref{fig: architectures}). 
First, hierarchical architecture arranges agents in a tree-like structure, allowing higher-level agents to delegate tasks to lower-level counterparts. 
For example, MDAgents~\cite{kim2024mdagents} introduces an approach where a Moderator managed multi-disciplinary clinicians. Depending on the different medical diagnosis requirements, the Moderator can recruit different clinicians, thereby improving the precision of disease treatment.
Next, decentralized architecture enables agents to operate autonomously and communicate directly. 
For instance, Chen et al.~\cite{chen2024scalable} propose a framework called DMAS, where every robot’s LLM agent can engage in task planning through turn-taking dialogues. This approach enables agents to express their views independently while integrating feedback from others, thereby enhancing collaboration for task completion.
Then, centralized architecture relies on a single authority to oversee and manage all agents. 
Magentic-one~\cite{fourney2024magentic} utilizes this type of communication architecture, where web agents, file agents, coders, and other agent workers all respond to the orchestration and direction of a central orchestrator. 
Finally, shared message pool architecture allows agents to interact through a common set of messages, promoting efficient information exchange.
An illustrative example is MetaGPT~\cite{hong2024metagpt}, which features a shared message pool that enables agents to continuously access and gather pertinent information, thus improving their collaborative efforts and overall communication efficiency.
In our work, we focus on studying the diagnosis of centralized multi-agent systems.

\subsection{Agentic System Diagnosis Tools}
Diagnosis of a system refers to identifying the location and cause of failures within that system~\cite{isermann2005fault}.
The insights gained from diagnosis can serve to improve the system in subsequent stages.
While a range of diagnosis tools exist for LLM-based single-agent systems, such as covering prompt-testing interfaces~\cite{epperson2025interactive,jiang2022promptmaker,tenney2024interactive} and output-comparison systems~\cite{kahng2024llm,kim2024evallm,shankar2024validates}, they often fail to address the broader challenges of multi-agent collaboration, which demand synchronized, system-wide analysis of multiple agents, their outputs, and their inter-communication~\cite{tran2025multi}.

Currently, platforms like AutoGen Studio~\cite{dibia2024autogen}, CrewAI~\cite{crewAI}, Voiceflow~\cite{voiceflow}, and LangGraph~\cite{LangchainAi}, use chat-style interfaces to help developers interact with AI agents.
These tools are capable of streaming the behaviors and generated messages (such as code snippets and created files) from the multi-agent system into the interface. 
Furthermore, several of them, like AutoGen Studio, also offer an overview interface that allows users to monitor metrics such as the number of times different agents are invoked and the amount of tokens consumed.
However, the overview provided by these tools often falls short in helping users identify the key points that lead to task failures. Instead, the focus tends to be on resource consumption issues, such as excessive execution time.
Additionally, this streamlined presentation of information can overwhelm developers, making it difficult to effectively understand the overall logic of task execution, thereby hindering the diagnosis of issues.
Other works, such as LangSmith~\cite{mavroudis2024langchain}, have introduced visualization methods for presenting the multi-agent execution process, which can enhance developers' ability to locate and understand failures. 
However, it primarily focuses on improving the visual appeal of the original execution sequence.
From those sequentially arranged execution details, developers still face challenges in effectively achieving a high-level understanding of the task execution, such as identifying frequent failure points and recognizing where multiple attempts were made.
Recently, AGDebugger~\cite{epperson2025interactive} has been proposed to assist developers in debugging LLM-based multi-agent systems; however, it mainly helps AI developers improve systems after failures have been identified and understood.

In response, there is currently a lack of effective tools to help users identify and understand failures in multi-agent behaviors.
Therefore, we propose a three-layered approach based on Activity Theory~\cite{kuutti1996activity} to organize agent self-actions and interactions, focusing on the diagnosis of centralized multi-agent systems.
Our system allows developers to switch between high-level behavior summaries and deeper log inspections as needed. 
By mapping how plans evolve, how agents coordinate tasks, and how errors compound over multiple rounds, developers can more efficiently understand failures.

\section{Formative Study}\label{sec:formative-study}
To identify pain points in locating and understanding failures for diagnosing multi-agent systems and to derive design requirements of our system, we conducted a qualitative formative study.

\subsection{Participants and Procedure}

We interviewed eight multi-agent project practitioners (E1–E8; 3 female, 5 male) recruited through social media and word-of-mouth referrals. 
The participants represented diverse roles and experience levels, including a startup founder (E1) who had led five multi-agent projects, two product developers (E2–E3), and five Ph.D. students specializing in AI or HCI (E4–E8).

During the interview, participants were asked to describe their most recent experience developing a multi-agent system. We prompted them to provide concrete examples of failures they encountered during development, along with the strategies they used to diagnose these issues. We then followed up with questions about the challenges they faced throughout the diagnostic process, encouraging them to elaborate with detailed examples. The full list of interview questions is provided in \autoref{formative:questions}. All interviews were audio-recorded. Each interview lasted approximately 90 minutes, and participants received US \$15 in compensation. This study was approved by the Institutional Review Board.

\rui{We analyzed the interview data using the thematic analysis proposed by Braun and Clarke~\cite{braun2006using}. 
All videos were transcribed, and then they were independently coded inductively by two researchers, focusing on participants’ diagnostic workflows and the challenges they encountered.
During open coding, the researchers code the participants' self-reported diagnostic activities with concrete, process-oriented codes (e.g., \textit{forming an initial overview}, \textit{reviewing agent logs}, and \textit{searching for anomalous behaviors}), with each code simultaneously annotated with the associated challenges.
After independent coding, the two researchers collaboratively discussed the codes to derive higher-level themes by clustering related codes.
As the first goal of this analysis was to identify common steps in the diagnosis process, codes were grouped into themes primarily based on similarities in diagnostic intent and temporal position within the diagnosis workflow.
When coding disagreements arose, the researchers revisited the original transcripts and engaged in iterative discussion to reconcile differing interpretations and reach consensus.
This process resulted in three major diagnostic steps.
The two researchers also reviewed and named these three steps.
Within each diagnostic step, the two researchers then applied a similar iterative clustering process to identify recurring challenges.
}
These themes form the basis of the findings reported in the following section, supported by representative participant quotes.

\subsection{Diagnostic Process and Challenges}
Through our qualitative analysis, we identified three common steps in multi-agent system diagnosis, along with the key challenges experts encounter during this process.

\textbf{\emph{Gaining an Overall Understanding.}}
A typical initial step of multi-agent system diagnosis involves making sense of the large volumes of diverse agent behaviors to gain an overall understanding of how they address the user query.
This overview can help users identify any inconsistencies in the overall task execution logic.
For instance, E3 discussed the experience with a stock analysis agentic system: \textit{``I needed to ensure that the system first gathered real-time news to assess market sentiment, then pulled historical prices to identify trends. Finally, the agentic system combined these elements to perform the analysis. Only if this workflow is correct do the specific behaviors at each step make sense.''}
However, all of our experts reported a high cognitive load during this understanding process.
They mentioned that they often need to carefully review information generated by agent operations (E1-E8), and take notes and reorganize the information (E2, E3, and E6) to make sense of the data, such as constructing their plan structures (E2, E6).
\textbf{This challenge intensified when different contexts required different types of summaries (E1, E2, E3, E6, and E8).} For example, E6 mentioned that she sometimes needed to quickly review updates across all plans, while at other times she needed an overview of the operations within a specific plan.
\textbf{Additionally, in some cases, our experts (E2, E4, and E6) even extract information not explicitly present in the logs} to enhance their understanding. 
For example, E2 often manually infers whether the planned actions and actual executions are consistent, since this information is not explicitly recorded in the log data.
This bottom-up sense-making process was thought of as time-consuming and mentally demanding.

\textbf{\emph{Locating Failures.}}
Then, developers will search for specific agent behaviors that reveal the root causes of failures. 
However, experts have highlighted the challenges in determining where to focus attention first (E2, E3, E4, E6, and E7) because \textbf{this often requires examining multiple interrelated agent behaviors buried within large volumes of logs}.
As E4 noted, \textit{``For example, if the behaviors of these agents are all aimed at completing the CT report analysis, it is important to consider them collectively. Analyzing a single agent's behavior in isolation may not provide meaningful insights; for instance, while one action may have failed, subsequent actions could potentially compensate for that failure.''}
In some cases, two agent behaviors may occur far apart in the log data, such as one appearing at the beginning and the other in the middle, yet still be very similar, and developers hope to analyze them together.
Unfortunately, the current method requires users to manually associate the execution step purpose with specific agent operations, which is quite arduous.

\textbf{\emph{Interpreting Failures with Context.}}
Identifying one failure does not mark the end of the process. AI developers often need to interpret these failures within the broader conversational context to understand how and why the breakdown occurred. 
However, they often \textbf{struggle to relate failures to the surrounding context within unstructured log data} to determine the best approach for fixing those failures (E1, E3, E4, E7, and E8).
For example, E4 and E7 mentioned that they may discover failures at different levels, such as some behaviors that were not supposed to execute at all since they could not achieve the task's objectives, while others were intended to execute but resulted in execution errors.
E7 stated, \textit{``Some failures may have higher priority, but I ultimately need to manually categorize them. When faced with a large number of identified failures, this process is often very time-consuming.''}

\subsection{Design Considerations}
Based on our findings, we have summarized four design considerations for diagnosing multi-agent systems.

\begin{enumerate}[label=\textbf{DC\arabic*}, leftmargin=*]
    \item \textit{\textbf{Provide multi-layer behavior summary.}}\label{DC1}
    We found that developers traditionally rely on large, unstructured log data to obtain an overview of multi-agent execution. This creates substantial cognitive load and highlights the need for behavior summaries. Moreover, several participants reported requiring different levels of summaries depending on the context. Therefore, the system should provide multi-layer summaries of agent behaviors, enabling users to flexibly select the appropriate overview on demand. 
    Such a multi-layer structure can also help developers integrate contextual information across different levels to better interpret identified failures. 
    
    \item \textit{\textbf{Elicit hidden thoughts of agent behaviors.}}\label{DC2}
    Our experts noted that when constructing an overview of the system’s execution, they often needed to extract information that was not explicitly provided in the logs or model outputs. However, performing these inferences across large numbers of behaviors is highly time-consuming and prone to error. Therefore, the system should automatically surface such implicit information. For instance, we can show behaviors that concretely contributed to each step, when and why a step was revised after a failure, and how each behavior aligns with the intended workflow. This reduces users’ cognitive burden and enables more precise diagnosis of coordination or reasoning issues.

    \item \textit{\textbf{Support grouping and joint analysis of different agent behaviors.}}\label{DC3}
    \rui{Experts emphasized that diagnosing failures often requires reasoning about sets of agent behaviors that are logically or functionally related, such as those contributing to the same subtask, pursuing the same intermediate goal, or compensating for earlier failures. 
    However, these relationships can be implicit and scattered across large volumes of logs, forcing developers to manually infer and assemble relevant behaviors, which is cognitively demanding and error-prone.}
    Therefore, our system should explicitly link such related agent behaviors, enabling developers to analyze them jointly.

    \item \textit{\textbf{Present the information generated by agents in a structured form.}}\label{DC4}
    \rui{Multi-agent conversational logs often contain heterogeneous information, such as their objectives, processes, outcomes, and updates. To facilitate comprehension and navigation, our system should present these diverse types of information in a clear and structured format, such as categorized tables or hierarchical lists.}
    Therefore, users can more efficiently navigate, filter, and relate information. This also helps them form a clear overview of the agent execution process and identify failures more effectively.
    
\end{enumerate}
\section{Framework}
To address the challenges in diagnosing multi-agent systems, we propose a framework for extracting, organizing, and presenting agent behaviors at multiple levels.
The framework is grounded in the progress of task completion, a design choice motivated by the diagnostic context: users care most about understanding why tasks fail and how to fix them.
This section introduces the framework designs and technically explains how the required information is extracted by probing a centralized multi-agent system.

\begin{figure}[ht]
\centering
\includegraphics[width=\linewidth]{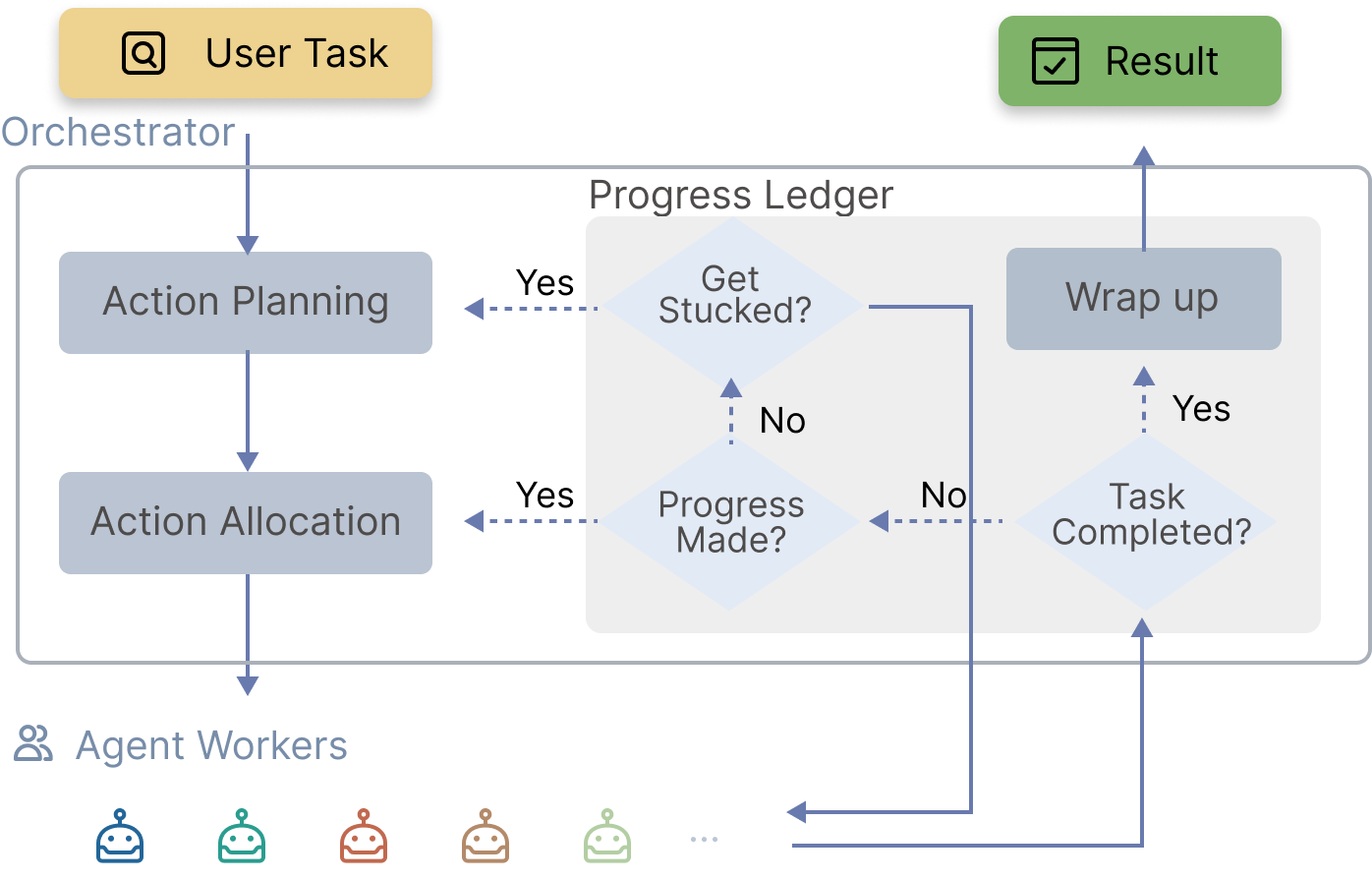}
\caption{The execution process in centralized LLM-based multi-agent systems.} 
\Description{This figure illustrates the execution process of a centralized LLM-based multi-agent system. A user task is first sent to a central orchestrator, which is responsible for overall coordination and control. The orchestrator performs action planning to decompose the task and then carries out action allocation to assign subtasks to multiple agent workers that execute them. During execution, a progress ledger continuously monitors the system state by checking whether progress has been made, whether the system is stuck, and whether the task has been completed. If no progress is detected or the system becomes stuck, feedback loops send the process back to action planning or action allocation for replanning or reassignment. Once the task is completed, the system enters a wrap-up stage and produces the final result.}
\label{fig: centralized}
\end{figure}

\subsection{Layered Summary of Agent Behaviors}\label{workflow}

From our formative study, we found that relying solely on plan log information makes it difficult to gain a holistic understanding of agent behaviors. To address this, we aim to develop a multi-layered summary that provides a structured yet comprehensive overview of the agents’ behaviors (\textbf{DC1}).

The key challenge lies in determining the layers of behaviors that carry concrete semantics and can highlight specific types of agentic issues. 
We were inspired by Activity Theory (AT), a framework for understanding human behaviors and practices within socially situated contexts~\cite{leont1974problem, Leontev1978ActivityCA}.
AT has been widely adopted in human–computer interaction research to analyze the interplay between humans and technology~\cite{kuutti1996activity}.

We adapt the framing of activities in AT to describe and organize agent behaviors as a way to complement user queries. Specifically, AT conceptualizes an activity as a purposeful, socially mediated system that can be understood at multiple levels: the \textbf{activity} itself, which is driven by a broader motive; \textbf{actions}, which are goal-directed processes that realize the activity; and \textbf{operations}, which are routinized or automated procedures conditioned by the immediate context. 
For example, organizing a community volunteer event can be seen as an activity: the goal is to help a local community, the actions include coordinating volunteers and arranging tasks, and the operations are the fine-grained steps, such as sending emails, booking venues, and managing schedules.
This hierarchical framing provides a principled lens for structuring and interpreting layers of agent behaviors, enabling us to map system-level intentions, task-level strategies, and low-level operations in ways that can reveal different types of agentic issues. This can help developers better make sense of agent behaviors and diagnose the systems.

At the activity level, we aim to provide users with a high-level understanding of the system’s approach to completing the activity, namely, \textit{how it complements the user’s query, executes the plan, and adapts as the plan evolves}. The activity-level summary enables users to assess the overall organization of how the query is being processed and the progress made toward completing the activity. If the organization appears unsatisfactory, users can infer that the system is not capable of handling the query and thus requires substantial changes. Conversely, if the high-level organization is adequate, users can then identify specific actions where the system repeatedly fails and drill down further to locate the causes of the issues.

At the action level, we aim to support users to delve deeper into the execution of a specific action by providing an overview of all the operations performed to complete this action, including the specific operations undertaken, their success or failure, and the results returned. 
This enables users to access the worker agent's operation decomposition capability and drill down to locate the erroneous operations.

At the operation level, we enable users to examine the concrete details of the system’s behavior, including the log history, agent tool usage, and web access processes.
By checking the details, users could confirm their findings about the issues.

\subsection{Probing Centralized Multi-agent Systems}

In centralized multi-agent systems, a central agent, known as the orchestrator, manages overall task progress (\autoref{fig: centralized}).
By maintaining a global view of the system, the orchestrator can optimize resource allocation, prevent conflicts, and ensure consistent outcomes, which makes centralized control a widely studied paradigm for critical tasks~\cite{fourney2024magentic, jiang2023llm, ning2023skeleton, qiao2024autoact, suzgun2024meta}.
The orchestrator typically decomposes a complex task into a sequence of steps, with each step referred to as an action (action planning). It tries to complete these actions one by one by invoking specialized agents (action allocation).
For example, completing a ``document reading'' action might involve first calling a web-searching agent to retrieve the literature, followed by a ``file-reading'' agent to parse and analyze its contents. The ordered sequence of such agent operations collectively fulfills the action, contributing to the overall progress of the task.
At the same time, the orchestrator needs to manage the overall progress of the task. For example, it determines the next steps by checking whether the task is completed, whether there has been no progress, or if it has entered a loop. Based on this assessment, the orchestrator decides whether to directly output results, update the planned actions itself, or continue invoking other agents.

In an activity-level summary, we aim to provide users with a high-level understanding of how the multi-agent system completes an entire activity. To achieve this, we present the plan as a unit, which includes information about plan failures, plan updates, and action progress overview in each plan. This summary helps developers quickly grasp how the multi-agent system executed the activity, enabling them to identify necessary actions that are significant and worth exploring further. The specific information extracted from agent behaviors includes the following:

\begin{itemize}

    \item \textit{\textbf{Plan failures and updating}}: We refer to a sequence of actions capable of completing an entire activity as a plan. In an activity, a multi-agent system often attempts various actions, which may occasionally lead to failures. This necessitates consideration of how to update the plan. At this layer, we provide information on instances of plan failure and their causes, as well as details on how to update the plan accordingly. This can help developers understand how the entire activity is performed more effectively.

    \item \textit{\textbf{Actions in each plan with their status}}: The final status (completed, failed, or not started) of each action within a plan is crucial for developers to understand where the process may be stuck. This information helps users narrow down which specific action to examine closely, facilitating more efficient diagnosis.
    
    \item \textit{\textbf{Agents for each action}}: Displaying which agents execute those actions provides additional context and information. This can help developers understand how an action is executed based on their expertise, without needing to delve into the internal workings of the action itself. For example, if an action involves reading a paper using a web agent, developers can intuitively understand that this action is completed through an online process. Conversely, if a file agent is used, it indicates that the reading is done locally. In certain cases, this distinction provides clarity, allowing developers to make an informed judgment about whether this action warrants further investigation. This enhances their ability to prioritize their efforts effectively.
    
    \item \textit{\textbf{Action progress}}: Multiple agents may interleave their contributions to an action, making it difficult for developers to quickly understand each agent’s participation, their coordination patterns, and the resulting outcomes. This makes it challenging to pinpoint the root of failures. To address this, we provide summaries that make these aspects explicit and easier to analyze. These summaries reveal how the system navigated the action, including the successes and failures that occurred.

\end{itemize}

\begin{figure*}[ht]
\centering
\includegraphics[width=\linewidth]{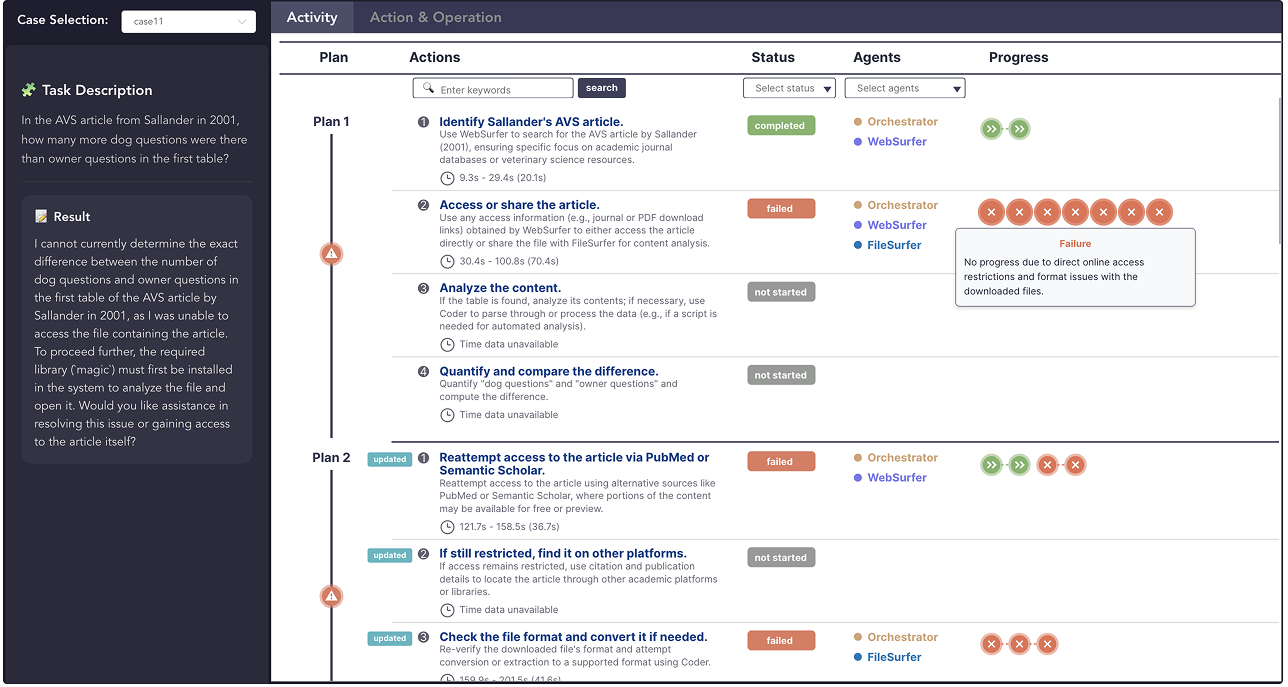}
\caption{The Activity View primarily helps AI developers gain a high-level understanding of the plans proposed by multi-agent systems and their execution status. Based on the user feedback of our formative study, we present five types of information at this level.} 
\Description{This figure shows the Activity View of a multi-agent system interface, which provides a high-level overview of task plans and their execution status. On the left, a sidebar displays the case selection, the user’s task description, and the final result or system message. The main panel on the right presents a structured timeline of multiple plans, each broken down into ordered actions. For each action, the interface shows its description, current status (such as completed, failed, or not started), the agents involved (for example, an orchestrator, web surfer, or file surfer), and visual indicators of progress. Failed actions are marked with warning icons and accompanied by brief failure explanations, such as access restrictions or file format issues. Multiple plans can be updated or retried when failures occur, allowing users to see how the system adapts its strategy over time.}
\label{fig: system1}
\end{figure*}

In an action-level summary, we enable users to delve deeper into the execution of a specific action after they identify which actions require further investigation. At this level, we provide an overview of all the operations performed to complete this action, including the specific operations undertaken, their success or failure, and the results returned. The detailed information includes the following:

\begin{itemize}
    \item \textit{\textbf{Operation description}}: An overview of each operation can illuminate the workflow involved in executing the action.
    This enables developers to quickly grasp how the action was executed.
    
    \item \textit{\textbf{Operation type}}: Providing a higher-level summary of operation descriptions can help developers effectively capture the relationships between operations in an action, aiding in identifying errors. For example, a web agent might have the capabilities to navigate URLs or read pages. If a developer discovers an issue with the ``read pages'' action, they can quickly validate the problem by examining the same or other similar types of actions.

    \item \textit{\textbf{Operation result summarization}}: The concise representation of the outputs can reveal critical insights into the overall effectiveness of each operation. By summarizing the results, developers can quickly assess whether the operations achieved the desired outcomes and how they contribute to the action's success or failure. This can help developers select which operations to drill down into for further investigation.
\end{itemize}

In operation-level details, we present the comprehensive log data for each operation, allowing developers to trace the original output of multi-agent systems.
\begin{itemize}

    \item \textit{\textbf{Log data}}: The concise representation of the outputs resulting from the operations performed by an agent.
    
\end{itemize}

To elicit agents' hidden thoughts for diagnostic purposes, such as plan formulation and revision, action status, and action progress, we employ natural language probing, using structured prompts that encourage self-evaluation and progress reporting (\hyperref[DC2]{\textbf{DC2})}. This probing enables us to capture key information across multiple levels of the system. In addition, we perform post-hoc processing to summarize operation goals and descriptions.
These prompts are adapted from the Magentic-One framework~\cite{fourney2024magentic}, a centralized multi-agent framework developed by Microsoft for solving complex tasks, which also extracts several types of information we need. The complete set of prompts is provided in the supplementary materials, where we have highlighted the modifications we made.
\section{DiLLS}

\begin{figure*}[h]
\centering
\includegraphics[width=\linewidth]{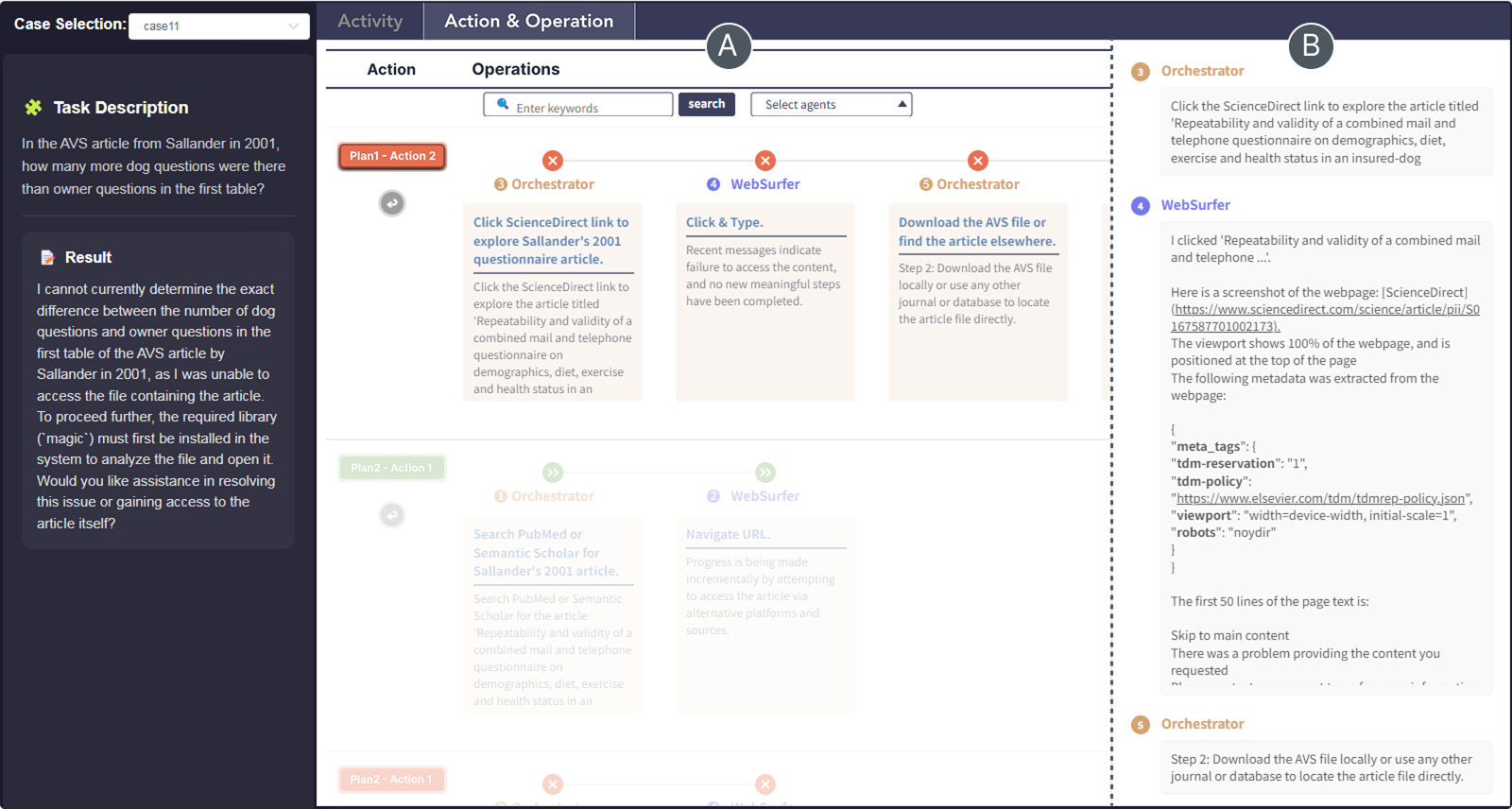}
\caption{The Action View and the Operation View support developers in accessing detailed information across two layers. The left side of this view (A) displays action-level information, while the right side (B) showcases operation-level information.} 
\Description{This figure presents the Action View and the Operation View of a multi-agent system interface, illustrating detailed execution information across two levels. On the left side (A), the interface shows action-level information organized by plans and actions, where each action is represented as a card labeled with the responsible agent, such as the orchestrator or a web surfer, and marked with its execution status, including failures or progress indicators. This view allows developers to trace how high-level plans are decomposed into concrete actions and how each action evolves over time. On the right side (B), the interface displays operation-level details corresponding to a selected action, including step-by-step logs, agent messages, accessed links, extracted metadata, and error messages. Together, the two sides enable developers to connect high-level action outcomes with low-level operational traces, helping them understand what the system attempted to do, how individual agents interacted with external resources, and why certain actions succeeded or failed.}
\label{fig: system2}
\end{figure*}

We introduce \tool, an interactive system that organizes task planning and execution information to support failure diagnosis.
\rui{A central design consideration is providing a multi-layered summary of agent behaviors (\hyperref[DC1]{\textbf{DC1}}). As discussed in Section~\ref{workflow}, agent behaviors are organized into three layers (i.e., activity, action, and operation) adapted from Activity Theory. \tool exposes these layers along the progression of task completion, enabling users to assess the overall activity, identify failed or stalled actions, and inspect operation-level details to verify root causes.}
Specifically, the Activity View offers a high-level overview of activity progress and outcomes, the Action View reveals action execution histories, and the Operation View displays raw operation-level logs.

\subsection{Activity View}\label{task}
The Activity View (\autoref{fig: system1}) provides AI developers with a high-level overview of activity execution, including overall progress and outcomes such as plans and their execution status.
\rui{This view corresponds to the first layer of our multi-layered design (\hyperref[DC1]{\textbf{DC1}}), enabling developers to quickly assess whether the activity is proceeding as intended before drilling down into finer-grained details.}
This view can be regarded as the first level of agent behavior summary, including five types of information which have been introduced in \autoref{workflow}  (\ie, plan failures and updating, actions in each plan, action status, agents for each action, and task progress).

\rui{Given the need to present multiple plans together with their associated action sequences, execution status, responsible agents, and progress updates, \tool~ adopts a table-based visualization (\autoref{fig: system1}). This design choice directly operationalizes \hyperref[DC4]{\textbf{DC4}} by structuring heterogeneous agent-generated information into clearly defined columns, enabling users to efficiently filter and navigate information while maintaining contextual relationships among plans, actions, and agents.}
Each type of information is displayed in a separate column, as shown in \autoref{fig: system1}.
The \textit{Plan column} shows consecutive plans and their execution status. 
A warning icon \raisebox{-.2\height}{\includegraphics[width=0.4cm]{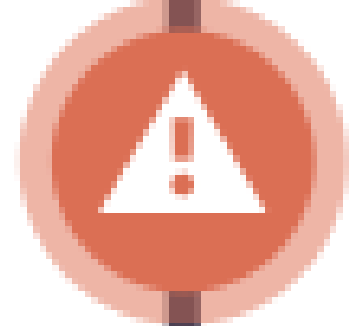}} between two plans allows users to quickly identify where the previous plan got stuck. 
Hovering over the icon, users can view the multi-agent system’s reasoning about the failure, such as possible causes and intended updates.
\rui{By making these thoughts explicit at the point of failure, users can more easily understand the rationale behind plan updates and examine whether the system’s transition from failure diagnosis to recovery strategy is well justified (\hyperref[DC2]{\textbf{DC2}}).}

When users seek more information about a particular plan, they can refer to the other columns to see how each plan is decomposed and executed.
For example, Plan 1 is broken down into four distinct actions, each represented as a separate row in the table. 
Specifically, four types of information are provided (i.e., descriptions, execution status, involved agents, and progress), each displayed in a separate column.
The \textit{Actions column} highlights a concise summary of the actions in this plan, with the execution time of each action also presented.
In addition, actions that are affected or updated in subsequent plans are prefixed with an update tag \raisebox{-.2\height}{\includegraphics[width=1cm]{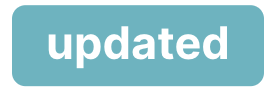}}, allowing users to quickly recognize dependencies and revisions across plans.
The \textit{Status column} then encodes whether an action is completed \raisebox{-.2\height}{\includegraphics[width=1cm]{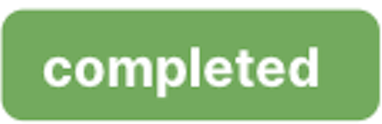}}, failed \raisebox{-.2\height}{\includegraphics[width=1cm]{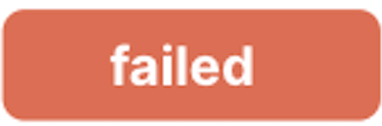}}, or not started \raisebox{-.2\height}{\includegraphics[width=1cm]{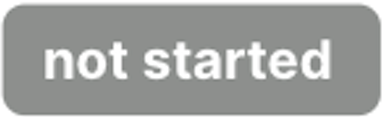}}.
In parallel, the \textit{Agents column} lists the agents involved in each action.

\rui{Finally, to make agents’ hidden assessments of task advancement interpretable to users, we explicitly externalize agents’ internal progress judgments (\hyperref[DC2]{\textbf{DC2}}). Through this, users can quickly assess whether and why progress was or was not achieved, and whether the agents’ own understanding of progress aligns with users’ expectations of task advancement.} Specifically, the \textit{Progress column} depicts this type of information.
Each marker represents the progress status of an operation within the action: making progress \raisebox{-.2\height}{\includegraphics[width=0.35cm]{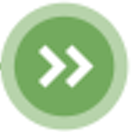}}, and no progress \raisebox{-.2\height}{\includegraphics[width=0.35cm]{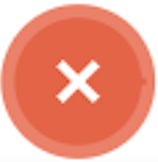}}.
\rui{
Considering that users often need to jointly reason about multiple agent behaviors rather than inspect them in isolation, we connect related operations (\hyperref[DC3]{\textbf{DC3}}).
Specifically, \tool segments progress into contiguous sequences based on progress signals, grouping consecutive operations that make progress and separating uninterrupted periods of no progress.
These segments, with concise summaries, act as coherent sense-making units, enabling users to identify not only when progress occurred or stalled, but also what was concretely accomplished.}
Clicking one segment navigates to the corresponding details in the \textbf{Action \& Operation View} (\autoref{Action and Log View}).
This design aims to let developers trace agent behaviors more effectively and to selectively narrow their focus during the exploration process.
\rui{Moreover, developers may need to jointly analyze agent behaviors that are not adjacent in the execution trace, such as operations performed by the same agent across different plans or operations pursuing the same intention across plan revisions.
To support this need, \tool provides flexible filtering mechanisms that allow users to regroup and examine related behaviors beyond their temporal order (\hyperref[DC3]{\textbf{DC3}}).}

\subsection{Action View}\label{Action and Log View}
\rui{The Action View (\autoref{fig: system2}-A) corresponds to the second layer of our multi-layered design (\hyperref[DC1]{\textbf{DC1}}), enabling users to quickly comprehend how the system progresses step by step toward completing an action.}
For example, when reading an online PDF, developers may want to know which agents were involved and what sequence of steps they performed. At this stage, users are often concerned only with whether these operations succeeded or failed, rather than the full content of each step. 
In this way, the Action View serves as an effective overview of progress. 
If needed, they can then drill down into the Operation View to inspect details such as the exact text retrieved by a web agent.

First, users can click on the progress marker from the Activity View to reach this view. 
Since the action level displays the sequence of operations within a single action, we adopted a list format (\autoref{fig: system2} - the left side) \hyperref[DC4]{(\textbf{DC4})}, allowing users to easily follow the progression of operations and understand the internal workflow of each action. Each list corresponds to a segment from the activity level. We continue to use markers to show progress and the corresponding agents, which maintains visual consistency.
Then, our system will display the operation description for each planning agent (\ie, the orchestrator) as it is responsible for issuing the execution commands.
Specifically, our system highlights the summary of each operation instruction, followed by a detailed description of this action.
This design choice aids in visual prioritization and draws attention to key details, allowing users to quickly identify important information. 
For the agent workers, we present their operation types in bold blue. Below that, we present the operation result summarization.
This format supports quick scanning, enabling users to rapidly assess different operation types and their outcomes, improving the efficiency of information sensemaking.

Finally, we have also added several interaction features to help users explore more effectively.
The system supports filtering interactions, similar to the Activity View. Users can search for corresponding operations using specific keywords or by selecting particular agents (\hyperref[DC3]{\textbf{DC3})}.
Additionally, to support navigation, when users click a progress segment in the Activity View, the corresponding list in the Action View will be highlighted.
At the same time, we increase the transparency of other unrelated lists.
This allows users to focus more on the content they care about.
Next, users can also click each return icon \raisebox{-.2\height}{\includegraphics[width=0.4cm]{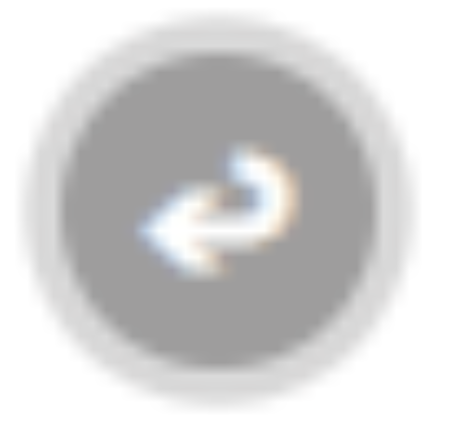}} to go back to the Activity View, and we use an animation to indicate which row in the Activity View corresponds to the user's operation.

\subsection{Operation View}
\rui{Corresponding to the third layer of our multi-layered design (\hyperref[DC1]{\textbf{DC1}}), the Operation View provides users with the ability to review the log details generated by an agent's operation (\autoref{fig: system2}-B).
In this view, we parsed the log data to produce structured outputs, making it easier for users to review (\hyperref[DC4]{\textbf{DC4}}).}
Furthermore, when users click on a row in the Action View, the relevant log details will be displayed in the Operation View.
In this way, users can easily navigate back and forth between different layers of information.
Finally, to make it easier for users to view the URLs provided by the WebSurfer, we have directly set them as hyperlinks.
\section{Evaluation}
\label{evaluation_design}
We conducted a within-subjects user study with 12 participants to evaluate the usability and effectiveness of our proposed system, \tool. 
According to the feedback from the formative study (Section \ref{sec:formative-study}), it is a common practice for developers to navigate the original log data using AutoGen Studio~\cite{dibia2024autogen}, CrewAI~\cite{crewAI}, or other tools to diagnose the behaviors of multi-agent systems. 
Therefore, in the control condition, participants were provided with observational log data.
To mitigate the influence of unrelated variables, this baseline system simply replaced the Activity View and the Action \& Operation View in \tool with a chronological presentation of log data, while retaining other components, such as the introduction and answers to the user query on the far left.
In the experiment condition, we provided participants with our interactive system, \tool, to help them explore the agent behaviors.

\rui{Findings from the formative study suggested that diagnosing LLM-based multi-agent systems often involves substantial effort and uncertainty, prompting us to investigate how our system affects efficiency, confidence, and cognitive load. Accordingly, we propose the following three research questions:}
(\textbf{RQ1}) whether our system enables developers to identify and understand failures more efficiently;
(\textbf{RQ2}) whether our system improves users’ confidence in diagnosing multi-agent systems; and
(\textbf{RQ3}) how users perceive the cognitive load of using our system, as well as their feedback on its functionality and design.

\rui{To operationalize these research questions, we evaluate diagnostic efficiency and effectiveness using task performance measures (e.g., the number and types of identified failures and inter-participant agreement), assess user confidence through post-task self-reported ratings, and examine cognitive load via NASA-TLX scores~\cite{hart2006nasa} and user experience via feature-level usability ratings. In addition, we conduct semi-structured interviews to collect qualitative insights that help contextualize these quantitative results.}

\subsection{Participants}
We recruited 12 participants (P1-P12) through online advertisements on social media and word-of-mouth.
The group included 6 females and 6 males, aged $24.17\pm 2.66$ years.
Participants came from both academia and industry.
They had diverse experiences in multi-agent system diagnosis: 12 had developed multi-agent systems for research purposes, and 3 had experience with projects in the industry.
Two of them participated in one multi-agent project, five contributed to two projects, three worked on three projects, and two engaged in more than five related projects.
Detailed demographic information of the participants is shown in \autoref{tab:participant_info}.

\begin{table}[ht]
  \centering
  \caption{Participant demographics, including each participant’s ID, age, specialization, and the number of multi-agent development projects they had previously participated in.}
  \Description{This table summarizes the demographics of twelve study participants, listing each participant’s ID, gender, age, area of specialization, and prior experience with multi-agent system development projects. The participants are labeled P1 through P12 and range in age from 21 to 30 years old. Most participants specialize in research, while a subset has experience spanning both industry and research. Both male and female participants are represented. The number of multi-agent development projects previously completed by participants varies from one to more than five, indicating a mix of relatively limited and more extensive prior experience within the group.}
  \label{tab:participant_info}
  \begin{tabular}{lcccc}
    \toprule
    \textbf{ID} & \textbf{Gender} & \textbf{Age} & \textbf{Specialization} &  \textbf{\#Projects}\\
    \midrule
    P1 & Male & 24 & Research & 2  \\
    P2 & Female & 25 & Industry \& Research & 3 \\
    P3 & Male & 24 & Research & >5 \\
    P4 & Female & 22 & Research & 2 \\
    P5 & Male & 21 & Research & 1 \\
    P6 & Female & 23 & Industry \& Research & 1 \\
    P7 & Female & 25 & Research & 3 \\
    P8 & Female & 21 & Research & 2  \\
    P9 & Male & 30 & Research & >5  \\
    P10 & Male & 23 & Research & 2 \\
    P11 & Male & 24 & Research & 2 \\
    P12 & Female & 28 & Industry \& Research & 3 \\
    \bottomrule
  \end{tabular}
\end{table} 

\subsection{Study Materials}\label{section:study-materials}
We utilized the GAIA benchmark~\cite{Mialon2024GAIA}, a commonly used dataset to evaluate multi-agent system performance, to identify suitable cases for our experiment.
For our multi-agent system, we used Magentic-One~\cite{fourney2024magentic}, one of the most widely used centralized multi-agent systems developed by Microsoft, which comprises five distinct agent roles. 
The Orchestrator manages the overall plans and executions, while the WebSurfer is responsible for tasks related to web searching. The FileSurfer performs operations associated with file handling, and the Coder engages in writing and debugging code. Finally, the Executor is tasked with executing the code.
To obtain suitable cases for the study, we tested 50 questions from the GAIA dataset. Ultimately, we selected four error cases that have a balanced level of difficulty, considering the number of agents, actions, and operations involved.
The first case involved the search question: \textit{``According to the US Bureau of Reclamation glossary, what does the acronym that shares its name with a book of the New Testament stand for''} (denoted as Case 1).
The second case asked: \textit{``In the AVS article from Sallander in 2001, how many more dog questions were there than owner questions in the first table''} (denoted as Case 2). 
The third case asked: \textit{``In the NIH translation of the original 1913 Michaelis-Menten Paper, what is the velocity of a reaction to four decimal places using the final equation in the paper based on the information for Reaction 7 in the Excel file''} (denoted as Case 3). 
The fourth case asked: \textit{``What is the average number of pre-2020 works on the open researcher and contributor identification pages of the people whose identification is in this file''} (denoted as Case 4).

\subsection{Procedure}
We first introduced the construction of Magentic-One~\cite{fourney2024magentic}, allowing participants to understand important information about the composition of agents within the framework, which serves as the foundation for subsequent system diagnosis. Then, we invited each participant to complete two multi-agent system diagnosis tasks separately under the two conditions. Before the task starts in the system condition, we provide a 5-minute tutorial on \tool and give 15 minutes for participants to get familiar with it. 
We prepared four candidate cases with wrong outputs (see details in Section \ref{section:study-materials}). Each participant completed two cases, one under the system condition and one under the baseline condition. To fully counterbalance which case was used in each condition, we generated all distinct pairs of two different cases (\eg Case 1 used in the system condition and Case 2 in the baseline condition, or the reverse).
The twelve resulting pairs were randomly assigned to the twelve participants. This ensured that every case appeared in both conditions across the study and that the case–condition assignments were fully counterbalanced.
In each task, participants were given 20 minutes to explore the case and identify the error using either our system or the baseline system. We encouraged them to think aloud during these two sessions, and we recorded the video of each user study session with participants' permission.
At the end of each session, participants were asked to explain their identified errors and propose potential solutions. Participants then completed a 5-point Likert questionnaire assessing the effectiveness, usability, and overall user experience of the tool. 
Upon the completion of the two sessions, we further conducted a semi-structured interview to gather participants' experiences and perceptions. 
Each participant received compensation of US\$20 for their time.
Our project has also received IRB approval.

\section{Results}
During the user study, our participants were encouraged to think aloud as they worked through the tasks. After each task, we conducted semi-structured interviews. The full list of questions is included in \autoref{userstudy:questions}. 
\rui{Specifically, the think-aloud data were used to inform the adaptation and refinement of follow-up interview questions to probe these observations in more depth. The qualitative analysis focused on the interview data.}
All interview sessions were video-recorded and transcribed.
\rui{Two authors independently coded the transcripts to generate initial codes, focusing on how the system’s features affected participants’ diagnostic efficiency, confidence, and cognitive load, each corresponding to one of the three research questions examined in \autoref{evaluation_design}.
These evaluation dimensions were examined separately, as they were elicited through different sets of interview questions.
Then, the two authors met to review and refine the codes, ultimately identifying representative codes and higher-level themes.
Specifically, during the process, they found that some themes might span multiple evaluation dimensions. For instance, clearer tracing of the exploration process was described by participants as contributing to both higher confidence and lower perceived cognitive load. When themes overlapped, we assigned emphasis based on the dimension that participants most strongly foregrounded in their accounts.}
These themes serve as qualitative findings that help contextualize the quantitative results, with participant quotations included as supporting evidence.

In this section, we first present the study results related to task performance. Next, we discuss user confidence in their diagnosis results, followed by an evaluation of their perceived load during the study. We then present additional important feedback provided by participants regarding our system features. Finally, we elaborate on the limitations of our user study.
Specifically, we found that the collected quantitative metrics from users did not follow a normal distribution. Therefore, we used the Wilcoxon signed-rank test to calculate significance.

\subsection{Task performance}
\tool helps users find more failures in the multi-agent system within a fixed time budget.
Specifically, under the baseline condition, each participant found an average of 2.25 (SD=1.23) failures. In contrast, under our system, participants identified an average of 3.67 (SD=1.84) failures. 
The results were statistically significant based on the Wilcoxon signed-rank test ($p < 0.05$), \rui{indicating that \tool is effective in assisting users in identifying failures, thus addressing \textbf{RQ1}.}
This is primarily because our system provides a hierarchical structure to organize agent behaviors. The different levels can help users identify various types of failures. Each structured layer highlights unsuccessful plans, actions, or operations, enabling our participants to quickly pinpoint agent behaviors that need detailed examination. Additionally, the transitions between layers allow users to easily recall the context.

To gain a deeper understanding of the identified failure points, we analyzed their types and examined the distribution of different types.
First, three authors independently annotated the failures reported by our participants using an open coding approach. After this initial annotation, the authors convened to discuss their classifications.
Ultimately, we established a final labeling schema. Each author then revised their previous annotations according to this schema. In cases where discrepancies arose, the authors engaged in discussions to determine the appropriate type for each failure.
Here are the four types of failures identified in our study:
\begin{itemize}
    \item \textit{Problematic Planning}: This indicates that the generated plans, consisting of specific actions, were impractical for fulfilling the user request or the updates to these plans were inappropriate.
    \item \textit{Action Skipping}: This refers to instances where the multi-agent system did not follow the predefined plan during its execution, potentially overlooking critical tasks.
    \item \textit{Incorrect Operation Assignment}: This means that the LLM-based multi-agent system did not give proper and clear instructions for the next operation or wrongly assigned operations to the agents responsible for completing them.
    \item \textit{Operation Completion Failure}: This indicates that the assigned operation was not successfully completed by the specific agent.
\end{itemize}

\begin{figure}[t]
\centering
\includegraphics[width=\linewidth]{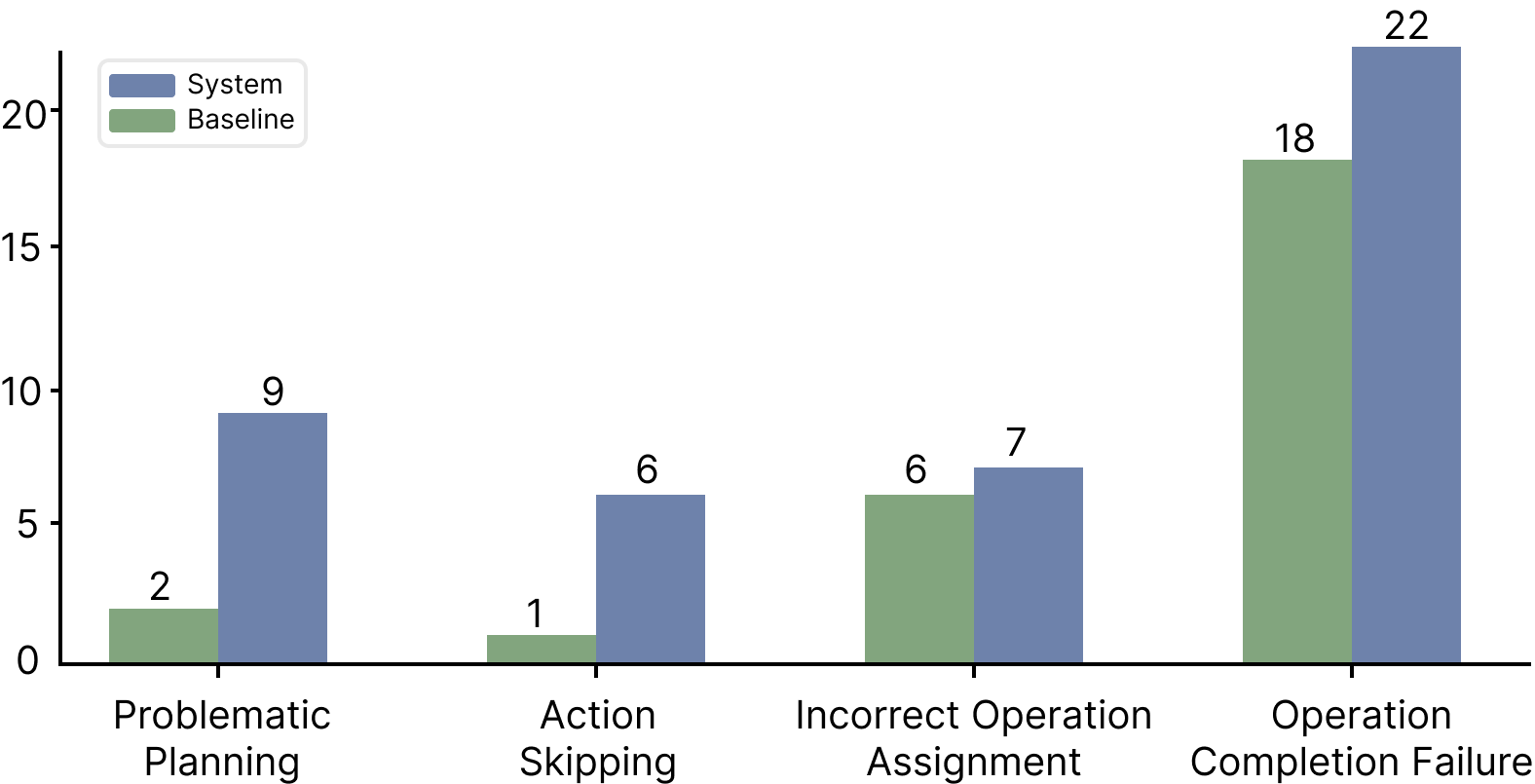}
\caption{The distribution of different types of failures.} 
\Description{This figure shows a bar chart comparing the distribution of different types of failures between a baseline system and the proposed system. The horizontal axis lists four failure categories: Problematic Planning, Action Skipping, Incorrect Operation Assignment, and Operation Completion Failure, while the vertical axis represents the number of observed failures. The specific number of failures can be checked in the text.}
\label{fig: failure}
\end{figure}

\autoref{fig: failure} shows the total number of failures identified across the four categories. \tool exhibits clear advantages in supporting users’ detection of these failures, particularly \textit{problematic planning} and \textit{action skipping}, both of which require a higher-level understanding of the overall execution process.

Additionally, we assessed inter-participant consistency in the identified failure types.
Due to the limited sample size, conventional, inter-rater agreement metrics (e.g., Fleiss’ Kappa) would not yield reliable estimates. 
Therefore, for the four examined cases, we instead report the proportion of failures identified by one, two, or all three participants (each case was independently reviewed by three participants under either the system or baseline condition).
In the system condition, 17\%, 25\%, and 58\% of failures were identified by one, two, and all three participants, respectively. 
In the baseline condition, the corresponding proportions were 50\%, 30\%, and 20\%. 
These results indicate substantially higher agreement among participants when using the system compared with the baseline, suggesting stronger alignment in diagnostic judgments.

Finally, we summarize two reasons why our system helps participants more easily detect the first two failure types.

\textit{Our system \textbf{strengthens information connections to support comparing actions across different plans.}}
Our activity-level summary enables seven participants to focus more on whether failures occur in the iterations between plans and the correspondence between plan design and action execution.
In the baseline, participants often became mired in operation details after understanding the plans, making it difficult to identify such issues in plan updates. 
For instance, P10 stated, \textit{``Plans and operations require constant back-and-forth scrolling for connection and comparison, so even when I want to check for action skipping or discrepancies, it is challenging and often inaccurate.''}
P4 also mentioned, \textit{``I believe identifying high-level failures related to overall execution logic is crucial, and I prefer to prioritize fixing these issues. If the execution logic is flawed, improving operational accuracy alone is unlikely to ensure task success. However, this can be quite challenging in practice. The two summary views effectively assist me in this process.''}
Additionally, our action-level summary serves a similar purpose. For instance, P4 noted that by summarizing all agent operations involved in completing an action, she can more easily identify differences and changes in the commands issued by the orchestrator.

\begin{figure*}[t]
\centering
\includegraphics[width=0.7\linewidth]{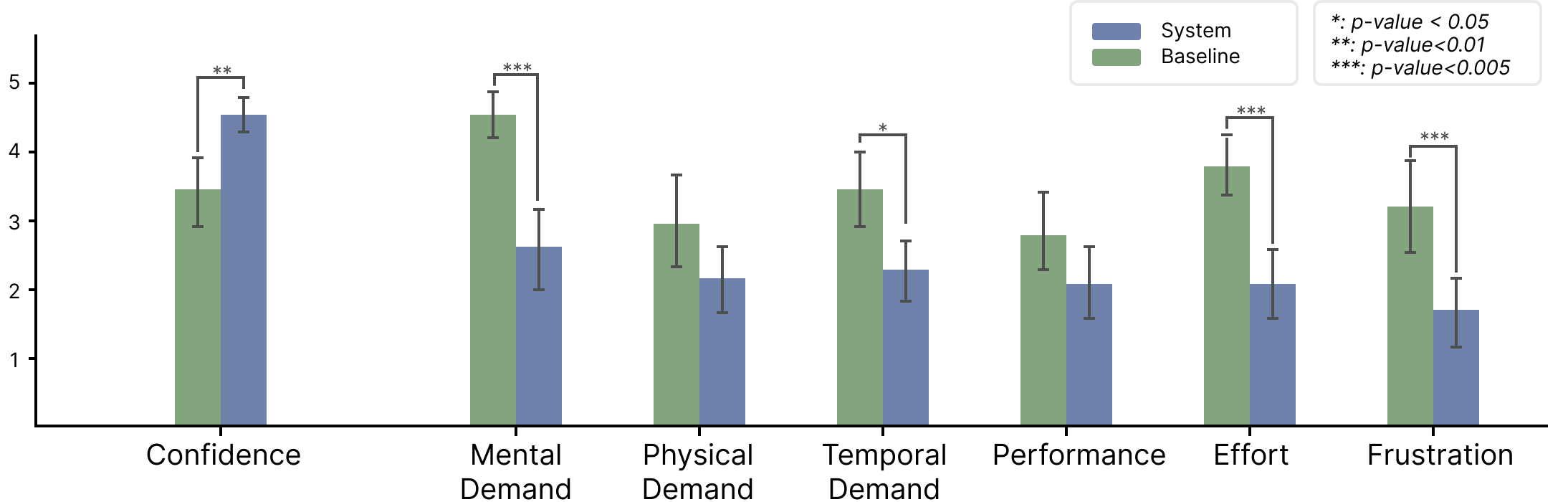}
\caption{User confidence and perceived load.} 
\Description{Two bar charts compare our system to the baseline. The left chart shows that our system significantly increases user confidence in identifying errors. The right chart displays NASA-TLX scores, showing that our system markedly reduces mental demand, effort, and frustration, proving the system's effectiveness in lowering the cognitive burden of diagnosis. The concrete rating scores can be checked in the text.}
\label{fig: confidence}
\end{figure*}

\textit{Our system \textbf{surfaces agents' internal reasoning to improve users’ understanding of their behaviors.}}
Five (of 12) participants mentioned that our system presents information they had not previously prompted the LLM-based agents to output, such as agents' perspectives on their own progress. This information helps them gain a deeper understanding of the reasons behind agents' executions, assisting in identifying potential failures.
For example, during P6's exploration, after the plan was updated and several operations executed, she did not see significant progress compared to the previous plan.
\textit{``In this updated plan, I observed that the agents still found the URL for the AVS article but failed to retrieve specific information, which is no different from the previous plan's execution.''} However, the agentic system considered it as progress. 
Curious, she checked the details and discovered that when the agents evaluated whether they made progress, they did not seem to take all the information obtained from the previous plans into account.
She stated, \textit{``This can be adjusted. We can enhance the agentic systems' understanding of all previous actions to help them evaluate progress more comprehensively. This can be achieved by implementing a global progress tracking mechanism.''}

\subsection{User Confidence}
\rui{We summarize participants’ self-reported confidence under the baseline and system conditions in \autoref{fig: confidence}. 
The results indicate that our participants reported higher confidence when using \tool, answering \textbf{RQ2}.} 
Specifically, under the baseline condition, each participant reported an average confidence level of 3.42 ($SD = 1.00$). In contrast, under our system, participants identified an average confidence level of 4.50 ($SD = 0.52$). 
The results were statistically significant based on the Wilcoxon signed-rank test ($p<0.05$), indicating that \tool is effective in enhancing user confidence in their identified failures and insights.
Based on the user feedback, we summarize the main reasons for participants having higher confidence in their identified failures when using \tool.

\textbf{\textit{Our system helps developers become aware of and trace their exploration process.}}
The two types of behavior summaries we provide are an effective way to help developers continuously understand their current exploration progress.
This allows users to have a clearer understanding of which parts they have explored and which they have not, reducing the likelihood of omissions. Ultimately, this leads to an increase in their confidence.
We observed that six participants frequently recalled their exploration processes in a structured way. 
For instance, P3 initially identified the action he was investigating. Then he utilized the progress summary to quickly gauge the contextual information after he started to examine the operation under this action.
He noted that the intuitive visual layout of each view aids him in tracing his exploration journey more effectively.
P10 also stated, \textit{``\tool appears to guide me through this diagnosis process using three layers, making me feel that my diagnostic approach has become more systematic and comprehensive.''}

\textit{\textbf{Our system highlights the statuses of plans, actions, and operations to help users reconfirm and refine their initial hypotheses, if any.}}
Six (of 12) participants highlighted the effectiveness of our system in emphasizing critical information, which plays a significant role in helping them confirm their initial hypotheses and even formulate more comprehensive hypotheses.
For example, P11 noted, \textit{``When I use the baseline, I often worry that I might not have considered all the hypotheses which need validation, which leads me to constantly check.''}
P3 also mentioned that, as a developer with extensive experience, they might quickly skim through all the log data and form their own hypotheses. However, they are concerned that their experience could introduce certain biases. The system directly points out the erroneous actions or operations, making it easier for them to confirm and even refine their thoughts.

\subsection{Perceived Load}
\rui{We summarize participants’ perceived cognitive load through NASA Task Load Index (NASA-TLX)~\cite{hart2006nasa} under the baseline and system conditions in \autoref{fig: confidence}.
We can observe that our system demonstrates advantages over the baseline across various dimensions, answering the first part of \textbf{RQ3}.}
We summarize two points to explain why our interactive system, \tool, is more advantageous than the baseline.

\textit{\textbf{Our system provides a wealth of hints to guide users, allowing them to bypass redundant in-depth examination.}}
Our activity-level summary outlines action statuses (i.e., completed, failed, and not started), effectively helping nine participants identify which actions to focus on. 
For instance, P10 mentioned, \textit{``In my previous work diagnosing multi-agent systems, the text volume was overwhelming, requiring me to scan everything to identify focus areas. This system highlights where the task is getting stuck.''}
Additionally, our progress summary also aids seven participants in quickly understanding the reasons for failures, which helps in formulating hypotheses about which operations to examine more closely.
For instance, P8 stated, \textit{``The text summary for operations under an action is very useful. It helps me identify which of the several operations have issues, allowing me to focus on those specific operations.''}
At the same time, seven (of 12) participants also mentioned that our action-level summary highlights the success and failure of operations, which aids their failure location process. P2 said, \textit{``This information, combined with the summary of what each operation did, allows me to quickly understand the entire execution trajectory and identify where further investigation is needed.''}

\textit{\textbf{The seamless transition between layers allows users to easily recall context.}}
Ten (of 12) participants appreciated the functionality that allows for seamless navigation and positioning between different levels.
This ability to examine context at any time significantly reduces their perceived load during the diagnostic process.
For instance, P7 noted that the transition animations are particularly smooth, and these animations help her understand where to focus her attention.
P8 mentioned that the use of the same encoding method between different layers, such as the icons for success and failure, provides visual consistency. This consistency has aided in her transition between layers.

\subsection{Further Analysis on User Experience Feedback} 

\begin{figure}[ht]
\centering
\includegraphics[width=\linewidth]{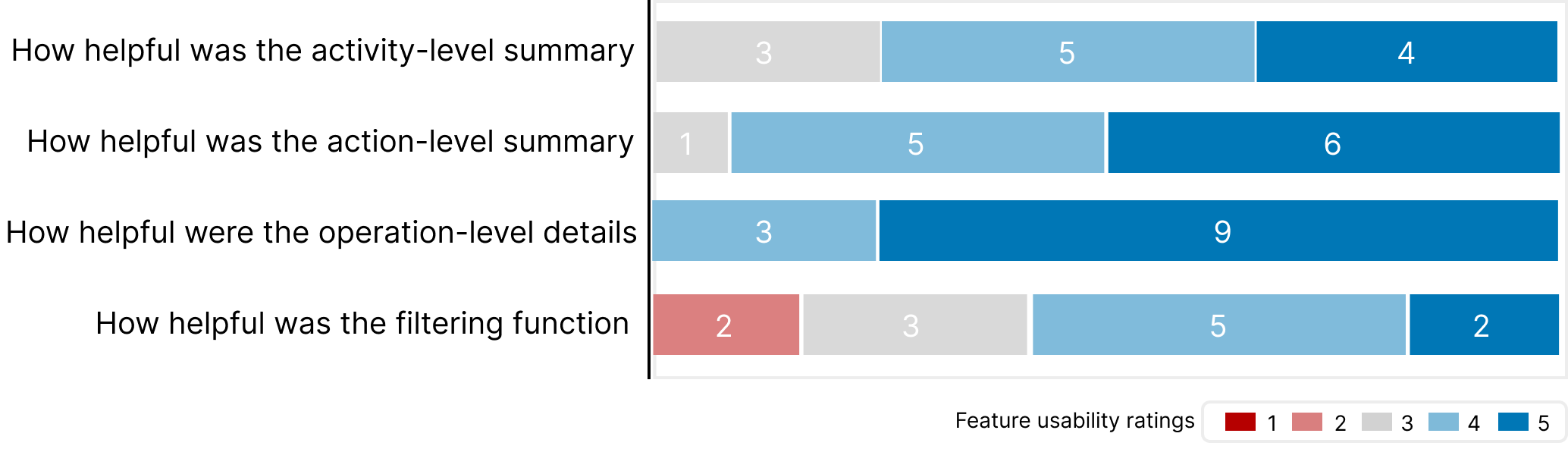}
\caption{User perception towards the system.} 
\Description{A horizontal stacked bar chart showing Likert-scale ratings for four system features. A detailed description of the score can be viewed in the text.}
\label{fig: usability}
\end{figure}

We analyzed user ratings of the usability for four features in our system: activity-level summary, action-level summary, operation-level details, and filtering function.
\rui{The distribution of the user ratings can be seen in \autoref{fig: usability}, addressing the second part of \textbf{RQ3}.}
We summarize two points explaining why our system achieves positive ratings through thematic analysis of the interview results.
Finally, we also elaborate on their concerns regarding our system.

\textit{\textbf{Our system's hierarchical organization of agent behaviors helps users notice failure types that they previously overlooked.}}
Four (of 12) participants appreciated that our system helped them identify failure types they had not previously considered.
P3 noted, \textit{``In the past, I never realized that task skipping could occur. However, this system has allowed me to efficiently detect when the agentic system starts the next action before the previous one has been executed.''}
P4 also mentioned, \textit{``These three layers provide a great taxonomy to help me identify and categorize failures. This systematic approach enables me to discover new types of failures.''}

\textit{\textbf{The filtering approach allows users to view successful and failed behaviors separately, enabling them to identify patterns and trends.}}
Five (of 12) participants utilized the filtering feature to extract actions with different statuses. Their goal was to gain a comprehensive understanding of which actions frequently failed and which were successful, enhancing their understanding of multi-agent capabilities and facilitating hypothesis generation. This exploration allowed them to form hypotheses about the model's performance and its underlying capabilities, which can help the diagnostic process.
P1 articulated this sentiment well, stating, \textit{``This is similar to global explainability for AI models; I want to quickly understand the boundaries of a model's capabilities—where it succeeds and where it fails.''}

\subsection{User Study Limitations}
\label{sec:user-study-limitation}

In this section, we mainly discuss some limitations of our user study. 
Our user study focuses solely on the initial step of debugging, helping users locate and understand failures by exploring agent behaviors (\ie, diagnosis). Therefore, after users generated hypotheses about how to improve multi-agent systems, we did not have them modify the system to test these hypotheses. This is primarily because actual debugging can be time-consuming, and we are unable to conduct this within the time constraints of our study.
Finally, since diagnosis can be closely related to each individual’s experience, opinions, skills, etc., it is challenging to provide a ground truth to evaluate whether the participants' results are correct. To alleviate this problem, we asked participants in the study to find corresponding evidence to support the failures they identified and to explain how they would like to address these failures.

\section{Discussion}
\label{sec: discussion}
In this section, we summarize several design implications for interactive tools that can support multi-agent system diagnosis and understanding, based on findings from our user study. We also discuss the generalizability and limitations of our work, along with directions for future research.

\subsection{Design Implications}
Our study highlights two key design implications for multi-agent system diagnosis: balancing AI assistance with human judgment and leveraging contextual information for failure analysis.

\smallskip

\textit{Balancing Automation and User Agency in Multi-agent System Diagnosis}.
Our system automates the extraction, structuring, and summarization of planning and execution traces, tasks that are tedious for humans, while leaving failure interpretation and pattern summarization to users. This keeps human judgment central and reduces the risk of misleading hallucinations from fully automated diagnoses. At the same time, automation highlights potentially relevant behaviors across large volumes of data, easing cognitive load and helping users navigate complex traces. This balance allows AI to amplify user expertise without replacing it. Future work may explore more proactive AI suggestions, such as proposing possible root failures with explanations, but these features must be designed so users can validate or override AI outputs to maintain interpretability, trust, and a balanced human–AI workflow.

\smallskip

\textit{Failure Analysis with Context}.
During our user study, we discovered that some participants frequently relied on successful actions or similar failures to assist them in analyzing current failures. This contextual information serves as a powerful resource for diagnosis. For example, users might identify a successful configuration that resolved a similar issue in the past or draw parallels between the characteristics of a current failure and those of prior cases.
Currently, our system facilitates context retrieval primarily by enabling users to easily access adjacent information or through basic filtering options. However, there is considerable room for improvement.
Future designs should not only enhance search capabilities but also focus on how the interface is structured to support users in grouping contextual information effectively.
Moreover, incorporating visual grouping tools or dashboards that display related actions and failures side by side could significantly aid users in their analysis. By creating a more intuitive interface that emphasizes context, we can empower users to make informed decisions based on historical insights, ultimately improving their diagnostic outcomes.

\subsection{Generalizability}

In this paper, we introduce a framework for extracting, organizing, and presenting task-planning and execution information to support the diagnosis of centralized multi-agent systems. The framework is conceptually grounded in activity theory, leveraging its three-layer structure to frame and interpret agentic activities. Because this structure captures fundamental and inherent properties of activity organization, we anticipate that the conceptual framework may generalize to other multi-agent architectures. However, the technical means required to obtain the relevant information may differ substantially across architectures.

In principle, task execution histories can be used to explicitly extract or summarize activity organization, action planning, and operational sequences. However, the difficulty of obtaining such information varies across multi-agent architectures. In systems that follow a plan-then-execute paradigm, commonly found in centralized or hierarchical architectures, there is typically an orchestrator agent~\cite{triedman2025multiagent,li2025disaster}, and its output history often provides direct access to planning and activity-organization information, enabling approaches similar to those used in this paper. In contrast, when planning and execution are interleaved, as in certain decentralized architectures or shared-message-pool systems ~\cite{li2025cooperative,yang2025agentnet}, obtaining high-level summaries becomes more challenging. In such cases, a bottom-up approach may be required, in which activity and action patterns are inferred from low-level traces, potentially supported by an observer agent that processes and aggregates the data. 

\subsection{Limitations and Future Work}

We have discussed the limitations of our user study in Section \ref{sec:user-study-limitation}. In addition, the use of LLMs for probing and summarizing text may also introduce inaccuracies. Although contemporary LLMs are generally robust for these tasks, errors can still occur. This risk can be mitigated, at least in part, through faithfulness evaluation methods~\cite{manakul2023selfcheckgpt}.
In the rest of this section, we outline opportunities for future work in understanding and diagnosing multi-agent systems, using our system and extending beyond it.

\smallskip

\smallskip

\textit{Understanding Long-Term Benefits.} 
Our user study is short-term and case-based. From a longer-term perspective, we aim to examine how the system influences practitioners’ diagnostic workflows. In particular, we seek to understand whether it can help practitioners develop more systematic and reusable diagnostic procedures. Moreover, we are interested in characterizing the patterns of model errors that emerge during extended diagnosis of a multi-agent system. Beyond insights tied to a single-case failure, we hope to uncover broader, systematic issues in system design, such as challenges in agent coordination or the lack of specific capabilities.

\smallskip

\textit{Broader Challenges in Understanding Multi-Agent Behaviors.}
Compared with single-agent systems, multi-agent systems introduce additional challenges for understanding their behaviors. In this paper, we focus primarily on their planning and execution behaviors as they work toward completing complex tasks. However, many other important questions remain open—for example, how organizational behaviors emerge, how agents’ roles evolve during problem solving~\cite{wang2025megaagent}, and what patterns arise from conflicting objectives and negotiation behaviors~\cite{kim2024mdagents}.
Together, these challenges highlight the growing complexity of behavior analysis as multi-agent systems become more capable and autonomous. We encourage future research to deepen the investigation of these emerging dynamics and develop methods that better support practitioners in understanding and guiding multi-agent systems at scale.

\smallskip

\textit{Toward A Theory-based Approach to Investigating Multi-agent System Behaviors.} 
Our use of Activity Theory (AT) focuses on its layered activity–action–operation structure to support practical diagnosis of multi-agent behaviors. 
However, AT offers far deeper analytical constructs, such as from aspects of mediation, community structures, rules, and contradictions that we do not explicitly model. 
These dimensions raise important questions about how agents coordinate, how tools and artifacts shape collective behavior, and how system-level tensions give rise to failures. While such inquiries extend beyond the scope of our diagnostic goals, situating our work within this broader AT landscape highlights promising directions for future research. We view \tool{} as an initial step toward more theory-informed analyses and hope it motivates deeper integration of AT concepts in the study of multi-agent system behaviors.

\section{Conclusion}
In this paper, we presented a framework and \tool, an interactive visualization system that facilitates the exploration of agent behaviors in LLM-based multi-agent systems for failure identification and understanding. Our system supports a layered exploration approach, grounded in Activity Theory, that enables comprehensive analysis of agent behaviors and allows AI developers to navigate activity-level, action-level, and operation-level information effectively. A user study with 12 AI developers demonstrates that our system can help practitioners identify and understand failures more efficiently and with greater confidence. Overall, our contributions provide a practical solution for diagnosing LLM-based multi-agent systems and pave the way for future research on developing more intuitive and effective tools for AI developers.
\section*{Declaration of the Use of AI}
We used ChatGPT for grammar checking and to improve word expressions. The tool did not change the original meaning of the text, nor did it introduce any new references or knowledge.

\begin{acks}
We sincerely thank all our collaborators for their contributions. We also appreciate the support provided by the Hong Kong government. This research was partly supported by the General Research Fund of the Research Grants Council (RGC GRF) of Hong Kong (Grant \#16210321), the Natural Science Foundation of Jiangsu Province (Grant \#BK20241300), and the Natural Sciences and Engineering Research Council of Canada (NSERC) Discovery (Grant \#RGPIN-2020-03966). 
In addition, we are extremely grateful to our reviewers for their careful reading of our manuscript and for providing detailed and constructive feedback.
\end{acks}

\bibliographystyle{ACM-Reference-Format}
\bibliography{main}


\begin{thebibliography}{60}


\ifx \showCODEN    \undefined \def \showCODEN     #1{\unskip}     \fi
\ifx \showISBNx    \undefined \def \showISBNx     #1{\unskip}     \fi
\ifx \showISBNxiii \undefined \def \showISBNxiii  #1{\unskip}     \fi
\ifx \showISSN     \undefined \def \showISSN      #1{\unskip}     \fi
\ifx \showLCCN     \undefined \def \showLCCN      #1{\unskip}     \fi
\ifx \shownote     \undefined \def \shownote      #1{#1}          \fi
\ifx \showarticletitle \undefined \def \showarticletitle #1{#1}   \fi
\ifx \showURL      \undefined \def \showURL       {\relax}        \fi
\providecommand\bibfield[2]{#2}
\providecommand\bibinfo[2]{#2}
\providecommand\natexlab[1]{#1}
\providecommand\showeprint[2][]{arXiv:#2}

\bibitem[Ala-Salmi et~al\mbox{.}(2025)]%
        {ala2025autonomous}
\bibfield{author}{\bibinfo{person}{Valtteri Ala-Salmi}, \bibinfo{person}{Zeeshan Rasheed}, \bibinfo{person}{Abdul~Malik Sami}, \bibinfo{person}{Zheying Zhang}, \bibinfo{person}{Kai-Kristian Kemell}, \bibinfo{person}{Jussi Rasku}, \bibinfo{person}{Shahbaz Siddeeq}, \bibinfo{person}{Mika Saari}, {and} \bibinfo{person}{Pekka Abrahamsson}.} \bibinfo{year}{2025}\natexlab{}.
\newblock \bibinfo{title}{Autonomous Legacy Web Application Upgrades Using a Multi-Agent System}.
\newblock
\showeprint[arxiv]{2501.19204}~[cs.SE]
\href{https://doi.org/10.48550/arXiv.2501.19204}{doi:\nolinkurl{10.48550/arXiv.2501.19204}}


\bibitem[Braun and Clarke(2006)]%
        {braun2006using}
\bibfield{author}{\bibinfo{person}{Virginia Braun} {and} \bibinfo{person}{Victoria Clarke}.} \bibinfo{year}{2006}\natexlab{}.
\newblock \showarticletitle{Using thematic analysis in psychology}.
\newblock \bibinfo{journal}{\emph{Qualitative research in psychology}} \bibinfo{volume}{3}, \bibinfo{number}{2} (\bibinfo{year}{2006}), \bibinfo{pages}{77--101}.
\newblock


\bibitem[Chen et~al\mbox{.}(2024b)]%
        {chen2023agentverse}
\bibfield{author}{\bibinfo{person}{Weize Chen}, \bibinfo{person}{Yusheng Su}, \bibinfo{person}{Jingwei Zuo}, \bibinfo{person}{Cheng Yang}, \bibinfo{person}{Chenfei Yuan}, \bibinfo{person}{Chi-Min Chan}, \bibinfo{person}{Heyang Yu}, \bibinfo{person}{Yaxi Lu}, \bibinfo{person}{Yi-Hsin Hung}, \bibinfo{person}{Chen Qian}, \bibinfo{person}{Yujia Qin}, \bibinfo{person}{Xin Cong}, \bibinfo{person}{Ruobing Xie}, \bibinfo{person}{Zhiyuan Liu}, \bibinfo{person}{Maosong Sun}, {and} \bibinfo{person}{Jie Zhou}.} \bibinfo{year}{2024}\natexlab{b}.
\newblock \showarticletitle{{AgentVerse: Facilitating Multi-Agent Collaboration and Exploring Emergent Behaviors}}. In \bibinfo{booktitle}{\emph{Proceedings of the Twelfth International Conference on Learning Representations}}, Vol.~\bibinfo{volume}{2024}. \bibinfo{publisher}{OpenReview.net}, \bibinfo{address}{Online}, \bibinfo{pages}{20094--20136}.
\newblock
\href{https://doi.org/10.48550/arXiv.2308.10848}{doi:\nolinkurl{10.48550/arXiv.2308.10848}}


\bibitem[Chen et~al\mbox{.}(2024a)]%
        {chen2024scalable}
\bibfield{author}{\bibinfo{person}{Yongchao Chen}, \bibinfo{person}{Jacob Arkin}, \bibinfo{person}{Yang Zhang}, \bibinfo{person}{Nicholas Roy}, {and} \bibinfo{person}{Chuchu Fan}.} \bibinfo{year}{2024}\natexlab{a}.
\newblock \showarticletitle{{Scalable Multi-Robot Collaboration with Large Language Models: Centralized or Decentralized Systems?}}. In \bibinfo{booktitle}{\emph{Proceedings of the 2024 IEEE International Conference on Robotics and Automation}}. \bibinfo{publisher}{IEEE}, \bibinfo{address}{Yokohama, Japan}, \bibinfo{pages}{4311--4317}.
\newblock
\href{https://doi.org/10.1109/ICRA57147.2024.10610676}{doi:\nolinkurl{10.1109/ICRA57147.2024.10610676}}


\bibitem[CrewAI(2024)]%
        {crewAI}
\bibfield{author}{\bibinfo{person}{CrewAI}.} \bibinfo{year}{2024}\natexlab{}.
\newblock \bibinfo{title}{{CrewAI: The Leading Multi-Agent Platform}}.
\newblock \bibinfo{howpublished}{\url{https://www.crewai.com/}}.
\newblock
\newblock
\shownote{Accessed: September 4, 2025}.


\bibitem[Dibia et~al\mbox{.}(2024)]%
        {dibia2024autogen}
\bibfield{author}{\bibinfo{person}{Victor Dibia}, \bibinfo{person}{Jingya Chen}, \bibinfo{person}{Gagan Bansal}, \bibinfo{person}{Suff Syed}, \bibinfo{person}{Adam Fourney}, \bibinfo{person}{Erkang Zhu}, \bibinfo{person}{Chi Wang}, {and} \bibinfo{person}{Saleema Amershi}.} \bibinfo{year}{2024}\natexlab{}.
\newblock \showarticletitle{{Autogen Studio: A No-code Developer Tool for Building and Debugging Multi-agent Systems}}. In \bibinfo{booktitle}{\emph{Proceedings of the 2024 Conference on Empirical Methods in Natural Language Processing: System Demonstrations}}. \bibinfo{publisher}{Association for Computational Linguistics}, \bibinfo{address}{Miami, Florida, USA}, \bibinfo{pages}{72--79}.
\newblock
\href{https://doi.org/10.18653/v1/2024.emnlp-demo.8}{doi:\nolinkurl{10.18653/v1/2024.emnlp-demo.8}}


\bibitem[Epperson et~al\mbox{.}(2025)]%
        {epperson2025interactive}
\bibfield{author}{\bibinfo{person}{Will Epperson}, \bibinfo{person}{Gagan Bansal}, \bibinfo{person}{Victor~C Dibia}, \bibinfo{person}{Adam Fourney}, \bibinfo{person}{Jack Gerrits}, \bibinfo{person}{Erkang~(Eric) Zhu}, {and} \bibinfo{person}{Saleema Amershi}.} \bibinfo{year}{2025}\natexlab{}.
\newblock \showarticletitle{{Interactive Debugging and Steering of Multi-Agent AI Systems}}. In \bibinfo{booktitle}{\emph{Proceedings of the 2025 CHI Conference on Human Factors in Computing Systems}}. \bibinfo{publisher}{ACM}, \bibinfo{address}{New York, NY, USA}, Article \bibinfo{articleno}{156}, \bibinfo{numpages}{15}~pages.
\newblock
\href{https://doi.org/10.1145/3706598.3713581}{doi:\nolinkurl{10.1145/3706598.3713581}}


\bibitem[Fan et~al\mbox{.}(2024)]%
        {fan2024contextcam}
\bibfield{author}{\bibinfo{person}{Xianzhe Fan}, \bibinfo{person}{Zihan Wu}, \bibinfo{person}{Chun Yu}, \bibinfo{person}{Fenggui Rao}, \bibinfo{person}{Weinan Shi}, {and} \bibinfo{person}{Teng Tu}.} \bibinfo{year}{2024}\natexlab{}.
\newblock \showarticletitle{{ContextCam: Bridging Context Awareness with Creative Human-AI Image Co-Creation}}. In \bibinfo{booktitle}{\emph{Proceedings of the 2024 CHI Conference on Human Factors in Computing Systems}}. \bibinfo{publisher}{ACM}, \bibinfo{address}{New York, NY, USA}, \bibinfo{pages}{1--17}.
\newblock
\href{https://doi.org/10.1145/3613904.3642129}{doi:\nolinkurl{10.1145/3613904.3642129}}


\bibitem[Fourney et~al\mbox{.}(2024)]%
        {fourney2024magentic}
\bibfield{author}{\bibinfo{person}{Adam Fourney}, \bibinfo{person}{Gagan Bansal}, \bibinfo{person}{Hussein Mozannar}, \bibinfo{person}{Cheng Tan}, \bibinfo{person}{Eduardo Salinas}, \bibinfo{person}{Friederike Niedtner}, \bibinfo{person}{Grace Proebsting}, \bibinfo{person}{Griffin Bassman}, \bibinfo{person}{Jack Gerrits}, \bibinfo{person}{Jacob Alber}, {et~al\mbox{.}}} \bibinfo{year}{2024}\natexlab{}.
\newblock \bibinfo{title}{Magentic-One: A Generalist Multi-Agent System for Solving Complex Tasks}.
\newblock
\showeprint[arxiv]{2411.04468}~[cs.AI]
\href{https://doi.org/10.48550/arXiv.2411.04468}{doi:\nolinkurl{10.48550/arXiv.2411.04468}}


\bibitem[Gao et~al\mbox{.}(2024)]%
        {GAO2024Empowering}
\bibfield{author}{\bibinfo{person}{Shanghua Gao}, \bibinfo{person}{Ada Fang}, \bibinfo{person}{Yepeng Huang}, \bibinfo{person}{Valentina Giunchiglia}, \bibinfo{person}{Ayush Noori}, \bibinfo{person}{Jonathan~Richard Schwarz}, \bibinfo{person}{Yasha Ektefaie}, \bibinfo{person}{Jovana Kondic}, {and} \bibinfo{person}{Marinka Zitnik}.} \bibinfo{year}{2024}\natexlab{}.
\newblock \showarticletitle{{Empowering Biomedical Discovery with AI Agents}}.
\newblock \bibinfo{journal}{\emph{Cell}} \bibinfo{volume}{187}, \bibinfo{number}{22} (\bibinfo{year}{2024}), \bibinfo{pages}{6125--6151}.
\newblock
\showISSN{0092-8674}
\href{https://doi.org/10.1016/j.cell.2024.09.022}{doi:\nolinkurl{10.1016/j.cell.2024.09.022}}


\bibitem[Ghafarollahi and Buehler(2024a)]%
        {GHAFAROLLAHI2024ProtAgents}
\bibfield{author}{\bibinfo{person}{Alireza Ghafarollahi} {and} \bibinfo{person}{Markus~J. Buehler}.} \bibinfo{year}{2024}\natexlab{a}.
\newblock \showarticletitle{{ProtAgents: Protein Discovery via Large Language Model Multi-agent Collaborations Combining Physics and Machine Learning}}.
\newblock \bibinfo{journal}{\emph{Digital Discovery}} \bibinfo{volume}{3}, \bibinfo{number}{7} (\bibinfo{year}{2024}), \bibinfo{pages}{1389--1409}.
\newblock
\href{https://doi.org/10.1039/d4dd00013g}{doi:\nolinkurl{10.1039/d4dd00013g}}


\bibitem[Ghafarollahi and Buehler(2024b)]%
        {Ghafarollahi2024SciAgents}
\bibfield{author}{\bibinfo{person}{Alireza Ghafarollahi} {and} \bibinfo{person}{Markus~J. Buehler}.} \bibinfo{year}{2024}\natexlab{b}.
\newblock \showarticletitle{{SciAgents: Automating Scientific Discovery Through Bioinspired Multi-Agent Intelligent Graph Reasoning}}.
\newblock \bibinfo{journal}{\emph{Advanced Materials}} \bibinfo{volume}{37}, \bibinfo{number}{22} (\bibinfo{year}{2024}), \bibinfo{pages}{2413523}.
\newblock
\href{https://doi.org/10.1002/adma.202413523}{doi:\nolinkurl{10.1002/adma.202413523}}


\bibitem[Hart(2006)]%
        {hart2006nasa}
\bibfield{author}{\bibinfo{person}{Sandra~G Hart}.} \bibinfo{year}{2006}\natexlab{}.
\newblock \showarticletitle{{NASA-task Load Index (NASA-TLX); 20 Years Later}}. In \bibinfo{booktitle}{\emph{Proceedings of the 2006 Human Factors and Ergonomics Society Annual Meeting}}. \bibinfo{publisher}{Sage Publications}, \bibinfo{address}{Los Angeles, CA, USA}, \bibinfo{pages}{904--908}.
\newblock
\href{https://doi.org/10.1177/154193120605000909}{doi:\nolinkurl{10.1177/154193120605000909}}


\bibitem[He et~al\mbox{.}(2024)]%
        {he2024llm}
\bibfield{author}{\bibinfo{person}{Junda He}, \bibinfo{person}{Christoph Treude}, {and} \bibinfo{person}{David Lo}.} \bibinfo{year}{2024}\natexlab{}.
\newblock \showarticletitle{{LLM-Based Multi-Agent Systems for Software Engineering: Literature Review, Vision and the Road Ahead}}.
\newblock \bibinfo{journal}{\emph{ACM Transactions on Software Engineering and Methodology}} \bibinfo{volume}{34}, \bibinfo{number}{5} (\bibinfo{year}{2024}), \bibinfo{pages}{1--30}.
\newblock
\href{https://doi.org/10.1145/3712003}{doi:\nolinkurl{10.1145/3712003}}


\bibitem[Hong et~al\mbox{.}(2024)]%
        {hong2024metagpt}
\bibfield{author}{\bibinfo{person}{Sirui Hong}, \bibinfo{person}{Mingchen Zhuge}, \bibinfo{person}{Jonathan Chen}, \bibinfo{person}{Xiawu Zheng}, \bibinfo{person}{Yuheng Cheng}, \bibinfo{person}{Ceyao Zhang}, \bibinfo{person}{Jinlin Wang}, \bibinfo{person}{Zili Wang}, \bibinfo{person}{Steven Ka~Shing Yau}, \bibinfo{person}{Zijuan Lin}, {et~al\mbox{.}}} \bibinfo{year}{2024}\natexlab{}.
\newblock \showarticletitle{{MetaGPT: Meta Programming for a Multi-Agent Collaborative Framework}}. In \bibinfo{booktitle}{\emph{Proceedings of the Twelfth International Conference on Learning Representations}}. \bibinfo{publisher}{OpenReview.net}, \bibinfo{address}{Online}, \bibinfo{pages}{1--20}.
\newblock
\href{https://doi.org/10.48550/arXiv.2401.12954}{doi:\nolinkurl{10.48550/arXiv.2401.12954}}


\bibitem[Huang et~al\mbox{.}(2023)]%
        {huang2023memory}
\bibfield{author}{\bibinfo{person}{Ziheng Huang}, \bibinfo{person}{Sebastian Gutierrez}, \bibinfo{person}{Hemanth Kamana}, {and} \bibinfo{person}{Stephen MacNeil}.} \bibinfo{year}{2023}\natexlab{}.
\newblock \showarticletitle{{Memory Sandbox: Transparent and Interactive Memory Management for Conversational Agents}}. In \bibinfo{booktitle}{\emph{Adjunct Proceedings of the 2023 36th Annual ACM Symposium on User Interface Software and Technology}}. \bibinfo{publisher}{ACM}, \bibinfo{address}{New York, NY, USA}, \bibinfo{pages}{1--3}.
\newblock
\href{https://doi.org/10.1145/3586182.3615796}{doi:\nolinkurl{10.1145/3586182.3615796}}


\bibitem[Hwang et~al\mbox{.}(2024)]%
        {hwang2024whose}
\bibfield{author}{\bibinfo{person}{Angel Hsing-Chi Hwang}, \bibinfo{person}{John~Oliver Siy}, \bibinfo{person}{Renee Shelby}, {and} \bibinfo{person}{Alison Lentz}.} \bibinfo{year}{2024}\natexlab{}.
\newblock \showarticletitle{{In Whose Voice?: Examining AI Agent Representation of People in Social Interaction through Generative Speech}}. In \bibinfo{booktitle}{\emph{Proceedings of the 2024 ACM Designing Interactive Systems Conference}}. \bibinfo{publisher}{ACM}, \bibinfo{address}{New York, NY, USA}, \bibinfo{pages}{224--245}.
\newblock
\href{https://doi.org/10.1145/3643834.3661555}{doi:\nolinkurl{10.1145/3643834.3661555}}


\bibitem[Isermann(2005)]%
        {isermann2005fault}
\bibfield{author}{\bibinfo{person}{Rolf Isermann}.} \bibinfo{year}{2005}\natexlab{}.
\newblock \bibinfo{booktitle}{\emph{{Fault-Diagnosis Systems: An Introduction from Fault Detection to Fault Tolerance}}}.
\newblock \bibinfo{publisher}{Springer}, \bibinfo{address}{Berlin, Germany}.
\newblock


\bibitem[Jiang et~al\mbox{.}(2023)]%
        {jiang2023llm}
\bibfield{author}{\bibinfo{person}{Dongfu Jiang}, \bibinfo{person}{Xiang Ren}, {and} \bibinfo{person}{Bill~Yuchen Lin}.} \bibinfo{year}{2023}\natexlab{}.
\newblock \showarticletitle{{LLM-Blender: Ensembling Large Language Models with Pairwise Ranking and Generative Fusion}}. In \bibinfo{booktitle}{\emph{Proceedings of the 61st Annual Meeting of the Association for Computational Linguistics: Volume 1: Long Papers}}. \bibinfo{publisher}{Association for Computational Linguistics}, \bibinfo{address}{Toronto, Canada}, \bibinfo{pages}{14165--14178}.
\newblock
\href{https://doi.org/10.18653/v1/2023.acl-long.792}{doi:\nolinkurl{10.18653/v1/2023.acl-long.792}}


\bibitem[Jiang et~al\mbox{.}(2022)]%
        {jiang2022promptmaker}
\bibfield{author}{\bibinfo{person}{Ellen Jiang}, \bibinfo{person}{Kristen Olson}, \bibinfo{person}{Edwin Toh}, \bibinfo{person}{Alejandra Molina}, \bibinfo{person}{Aaron Donsbach}, \bibinfo{person}{Michael Terry}, {and} \bibinfo{person}{Carrie~J Cai}.} \bibinfo{year}{2022}\natexlab{}.
\newblock \showarticletitle{{PromptMaker: Prompt-Based Prototyping with Large Language Models}}. In \bibinfo{booktitle}{\emph{Extended Abstracts of the 2022 CHI Conference on Human Factors in Computing Systems}}. \bibinfo{publisher}{ACM}, \bibinfo{address}{New York, NY, USA}, Article \bibinfo{articleno}{35}, \bibinfo{numpages}{8}~pages.
\newblock
\showISBNx{9781450391566}
\href{https://doi.org/10.1145/3491101.3503564}{doi:\nolinkurl{10.1145/3491101.3503564}}


\bibitem[Jin et~al\mbox{.}(2024)]%
        {jin2024teach}
\bibfield{author}{\bibinfo{person}{Hyoungwook Jin}, \bibinfo{person}{Seonghee Lee}, \bibinfo{person}{Hyungyu Shin}, {and} \bibinfo{person}{Juho Kim}.} \bibinfo{year}{2024}\natexlab{}.
\newblock \showarticletitle{{Teach AI How to Code: Using Large Language Models as Teachable Agents for Programming Education}}. In \bibinfo{booktitle}{\emph{Proceedings of the 2024 CHI Conference on Human Factors in Computing Systems}}. \bibinfo{publisher}{ACM}, \bibinfo{address}{New York, NY, USA}, \bibinfo{pages}{1--28}.
\newblock
\href{https://doi.org/10.1145/3613904.3642349}{doi:\nolinkurl{10.1145/3613904.3642349}}


\bibitem[Kahng et~al\mbox{.}(2024)]%
        {kahng2024llm}
\bibfield{author}{\bibinfo{person}{Minsuk Kahng}, \bibinfo{person}{Ian Tenney}, \bibinfo{person}{Mahima Pushkarna}, \bibinfo{person}{Michael~Xieyang Liu}, \bibinfo{person}{James Wexler}, \bibinfo{person}{Emily Reif}, \bibinfo{person}{Krystal Kallarackal}, \bibinfo{person}{Minsuk Chang}, \bibinfo{person}{Michael Terry}, {and} \bibinfo{person}{Lucas Dixon}.} \bibinfo{year}{2024}\natexlab{}.
\newblock \showarticletitle{{LLM Comparator: Interactive Analysis of Side-by-Side Evaluation of Large Language Models}}.
\newblock \bibinfo{journal}{\emph{IEEE Transactions on Visualization and Computer Graphics}} \bibinfo{volume}{31}, \bibinfo{number}{1} (\bibinfo{year}{2024}), \bibinfo{pages}{503--513}.
\newblock
\href{https://doi.org/10.1109/TVCG.2024.3456354}{doi:\nolinkurl{10.1109/TVCG.2024.3456354}}


\bibitem[Kim et~al\mbox{.}(2024a)]%
        {kim2024evallm}
\bibfield{author}{\bibinfo{person}{Tae~Soo Kim}, \bibinfo{person}{Yoonjoo Lee}, \bibinfo{person}{Jamin Shin}, \bibinfo{person}{Young-Ho Kim}, {and} \bibinfo{person}{Juho Kim}.} \bibinfo{year}{2024}\natexlab{a}.
\newblock \showarticletitle{{EvalLM: Interactive Evaluation of Large Language Model Prompts on User-Defined Criteria}}. In \bibinfo{booktitle}{\emph{Proceedings of the 2024 CHI Conference on Human Factors in Computing Systems}}. \bibinfo{publisher}{ACM}, \bibinfo{address}{New York, NY, USA}, Article \bibinfo{articleno}{306}, \bibinfo{numpages}{21}~pages.
\newblock
\href{https://doi.org/10.1145/3613904.3642216}{doi:\nolinkurl{10.1145/3613904.3642216}}


\bibitem[Kim et~al\mbox{.}(2024b)]%
        {kim2024mdagents}
\bibfield{author}{\bibinfo{person}{Yubin Kim}, \bibinfo{person}{Chanwoo Park}, \bibinfo{person}{Hyewon Jeong}, \bibinfo{person}{Yik~S Chan}, \bibinfo{person}{Xuhai Xu}, \bibinfo{person}{Daniel McDuff}, \bibinfo{person}{Hyeonhoon Lee}, \bibinfo{person}{Marzyeh Ghassemi}, \bibinfo{person}{Cynthia Breazeal}, {and} \bibinfo{person}{Hae~W Park}.} \bibinfo{year}{2024}\natexlab{b}.
\newblock \showarticletitle{{MDAgents: An Adaptive Collaboration of LLMs for Medical Decision-Making}}. In \bibinfo{booktitle}{\emph{Proceedings of the Thirty-Seventh Conference on Neural Information Processing Systems}}, Vol.~\bibinfo{volume}{37}. \bibinfo{publisher}{Curran Associates, Inc.}, \bibinfo{address}{Vancouver, Canada}, \bibinfo{pages}{79410--79452}.
\newblock
\href{https://doi.org/10.5555/3737916.3740438}{doi:\nolinkurl{10.5555/3737916.3740438}}


\bibitem[Kuutti(1996)]%
        {kuutti1996activity}
\bibfield{author}{\bibinfo{person}{Kari Kuutti}.} \bibinfo{year}{1996}\natexlab{}.
\newblock \bibinfo{booktitle}{\emph{{Activity Theory as a Potential Framework for Human-Computer Interaction Research}}}.
\newblock \bibinfo{publisher}{MIT Press}, \bibinfo{address}{Cambridge, MA, USA}, \bibinfo{pages}{17}.
\newblock
\href{https://doi.org/10.7551/mitpress/2137.001.0001}{doi:\nolinkurl{10.7551/mitpress/2137.001.0001}}


\bibitem[Langchain-Ai(2024)]%
        {LangchainAi}
\bibfield{author}{\bibinfo{person}{Langchain-Ai}.} \bibinfo{year}{2024}\natexlab{}.
\newblock \bibinfo{title}{{Langchain-ai/Langgraph-studio: Desktop App for Prototyping and Debugging LangGraph Applications Locally}}.
\newblock
\urldef\tempurl%
\url{https://github.com/langchain-ai/langgraph-studio}
\showURL{%
\tempurl}


\bibitem[Leont'ev(1974)]%
        {leont1974problem}
\bibfield{author}{\bibinfo{person}{Aleksei~N Leont'ev}.} \bibinfo{year}{1974}\natexlab{}.
\newblock \showarticletitle{{The Problem of Activity in Psychology}}.
\newblock \bibinfo{journal}{\emph{Soviet psychology}} \bibinfo{volume}{13}, \bibinfo{number}{2} (\bibinfo{year}{1974}), \bibinfo{pages}{4--33}.
\newblock
\href{https://doi.org/10.2753/RPO1061-040513024}{doi:\nolinkurl{10.2753/RPO1061-040513024}}


\bibitem[Leont'ev and Hall(1978)]%
        {Leontev1978ActivityCA}
\bibfield{author}{\bibinfo{person}{A.~N. Leont'ev} {and} \bibinfo{person}{M~J Hall}.} \bibinfo{year}{1978}\natexlab{}.
\newblock \bibinfo{booktitle}{\emph{{Activity, Consciousness, and Personality}}}.
\newblock \bibinfo{publisher}{Prentice-Hall}, \bibinfo{address}{Englewood Cliffs, NJ, USA}. 186 pages.
\newblock
\urldef\tempurl%
\url{https://api.semanticscholar.org/CorpusID:141375009}
\showURL{%
\tempurl}


\bibitem[Li et~al\mbox{.}(2025b)]%
        {li2025disaster}
\bibfield{author}{\bibinfo{person}{Bo Li}, \bibinfo{person}{Junwei Ma}, \bibinfo{person}{Kai Yin}, \bibinfo{person}{Yiming Xiao}, \bibinfo{person}{Chia-Wei Hsu}, {and} \bibinfo{person}{Ali Mostafavi}.} \bibinfo{year}{2025}\natexlab{b}.
\newblock \bibinfo{title}{Disaster Management in the Era of Agentic AI Systems: A Vision for Collective Human-Machine Intelligence for Augmented Resilience}.
\newblock
\showeprint[arxiv]{2510.16034}~[cs.MA]
\href{https://doi.org/10.48550/arXiv.2510.16034}{doi:\nolinkurl{10.48550/arXiv.2510.16034}}


\bibitem[Li et~al\mbox{.}(2025a)]%
        {li2025new}
\bibfield{author}{\bibinfo{person}{Fan Li}, \bibinfo{person}{Su Han}, \bibinfo{person}{Ching-Hung Lee}, \bibinfo{person}{Shanshan Feng}, \bibinfo{person}{Zhuoxuan Jiang}, {and} \bibinfo{person}{Zhu Sun}.} \bibinfo{year}{2025}\natexlab{a}.
\newblock \showarticletitle{{A New Era in Human Factors Engineering: A Survey of the Applications and Prospects of Large Multimodal Models}}.
\newblock \bibinfo{journal}{\emph{International Journal of Human--Computer Interaction}} \bibinfo{volume}{41}, \bibinfo{number}{18} (\bibinfo{year}{2025}), \bibinfo{pages}{1--14}.
\newblock
\href{https://doi.org/10.1080/10447318.2024.2446511}{doi:\nolinkurl{10.1080/10447318.2024.2446511}}


\bibitem[Li et~al\mbox{.}(2023)]%
        {li2023camel}
\bibfield{author}{\bibinfo{person}{Guohao Li}, \bibinfo{person}{Hasan Hammoud}, \bibinfo{person}{Hani Itani}, \bibinfo{person}{Dmitrii Khizbullin}, {and} \bibinfo{person}{Bernard Ghanem}.} \bibinfo{year}{2023}\natexlab{}.
\newblock \showarticletitle{{Camel: Communicative Agents for Mind Exploration of Large Language Model Society}}. In \bibinfo{booktitle}{\emph{Proceedings of the Thirty-Sixth Conference on Neural Information Processing Systems}}, Vol.~\bibinfo{volume}{36}. \bibinfo{publisher}{Curran Associates, Inc.}, \bibinfo{address}{New Orleans, LA, USA}, \bibinfo{pages}{51991--52008}.
\newblock
\href{https://doi.org/10.5555/3666122.3668386}{doi:\nolinkurl{10.5555/3666122.3668386}}


\bibitem[Li et~al\mbox{.}(2024)]%
        {li2024survey}
\bibfield{author}{\bibinfo{person}{Xinyi Li}, \bibinfo{person}{Sai Wang}, \bibinfo{person}{Siqi Zeng}, \bibinfo{person}{Yu Wu}, {and} \bibinfo{person}{Yi Yang}.} \bibinfo{year}{2024}\natexlab{}.
\newblock \showarticletitle{{A Survey on LLM-based Multi-Agent Systems: Workflow, Infrastructure, and Challenges}}.
\newblock \bibinfo{journal}{\emph{Vicinage Earth}} \bibinfo{volume}{1}, \bibinfo{number}{1} (\bibinfo{year}{2024}), \bibinfo{pages}{9}.
\newblock
\href{https://doi.org/10.1007/s44336-024-00009-2}{doi:\nolinkurl{10.1007/s44336-024-00009-2}}


\bibitem[Li et~al\mbox{.}(2025c)]%
        {li2025cooperative}
\bibfield{author}{\bibinfo{person}{Zhiyuan Li}, \bibinfo{person}{Wenshuai Zhao}, {and} \bibinfo{person}{Joni Pajarinen}.} \bibinfo{year}{2025}\natexlab{c}.
\newblock \bibinfo{title}{Cooperative Multi-Agent Planning with Adaptive Skill Synthesis}.
\newblock
\showeprint[arxiv]{2502.10148}~[cs.AI]
\href{https://doi.org/10.48550/arXiv.2502.10148}{doi:\nolinkurl{10.48550/arXiv.2502.10148}}


\bibitem[Lin et~al\mbox{.}(2025)]%
        {Lin2025survey}
\bibfield{author}{\bibinfo{person}{Yanna Lin}, \bibinfo{person}{Shaojie Xu}, \bibinfo{person}{Wenshuo Zhang}, \bibinfo{person}{Yushi Sun}, \bibinfo{person}{Zixin Chen}, \bibinfo{person}{Yanjie Zhang}, {and} \bibinfo{person}{Rui Sheng}.} \bibinfo{year}{2025}\natexlab{}.
\newblock \showarticletitle{A Survey of LLM-based Multi-agent Systems in Medicine}.
\newblock \bibinfo{journal}{\emph{Authorea Preprints}} (\bibinfo{year}{2025}).
\newblock
\href{https://doi.org/10.36227/techrxiv.176089343.36199495/v1}{doi:\nolinkurl{10.36227/techrxiv.176089343.36199495/v1}}


\bibitem[Liu et~al\mbox{.}(2024)]%
        {liu2024classmeta}
\bibfield{author}{\bibinfo{person}{Ziyi Liu}, \bibinfo{person}{Zhengzhe Zhu}, \bibinfo{person}{Lijun Zhu}, \bibinfo{person}{Enze Jiang}, \bibinfo{person}{Xiyun Hu}, \bibinfo{person}{Kylie~A Peppler}, {and} \bibinfo{person}{Karthik Ramani}.} \bibinfo{year}{2024}\natexlab{}.
\newblock \showarticletitle{{Classmeta: Designing Interactive Virtual Classmate to Promote VR Classroom Participation}}. In \bibinfo{booktitle}{\emph{Proceedings of the 2024 CHI Conference on Human Factors in Computing Systems}}. \bibinfo{publisher}{ACM}, \bibinfo{address}{New York, NY, USA}, \bibinfo{pages}{1--17}.
\newblock
\href{https://doi.org/10.1145/3613904.3642947}{doi:\nolinkurl{10.1145/3613904.3642947}}


\bibitem[Manakul et~al\mbox{.}(2023)]%
        {manakul2023selfcheckgpt}
\bibfield{author}{\bibinfo{person}{Potsawee Manakul}, \bibinfo{person}{Adian Liusie}, {and} \bibinfo{person}{Mark Gales}.} \bibinfo{year}{2023}\natexlab{}.
\newblock \showarticletitle{{SelfCheckGPT: Zero-Resource Black-Box Hallucination Detection for Generative Large Language Models}}. In \bibinfo{booktitle}{\emph{Proceedings of the 2023 Conference on Empirical Methods in Natural Language Processing}}. \bibinfo{publisher}{Association for Computational Linguistics}, \bibinfo{address}{Singapore}, \bibinfo{pages}{9004--9017}.
\newblock
\href{https://doi.org/10.18653/v1/2023.emnlp-main.557}{doi:\nolinkurl{10.18653/v1/2023.emnlp-main.557}}


\bibitem[Manish(2024)]%
        {manish2024autonomous}
\bibfield{author}{\bibinfo{person}{Sanwal Manish}.} \bibinfo{year}{2024}\natexlab{}.
\newblock \showarticletitle{{An Autonomous Multi-agent LLM Framework for Agile Software Development}}.
\newblock \bibinfo{journal}{\emph{International Journal of Trend in Scientific Research and Development}} \bibinfo{volume}{8}, \bibinfo{number}{5} (\bibinfo{year}{2024}), \bibinfo{pages}{892--898}.
\newblock
\href{https://doi.org/10.5281/zenodo.14179129}{doi:\nolinkurl{10.5281/zenodo.14179129}}


\bibitem[Mavroudis and Contributors(2024)]%
        {mavroudis2024langchain}
\bibfield{author}{\bibinfo{person}{Vasilios Mavroudis} {and} \bibinfo{person}{LangChain Contributors}.} \bibinfo{year}{2024}\natexlab{}.
\newblock \bibinfo{title}{LangChain v0.3}.
\newblock \bibinfo{howpublished}{\url{https://python.langchain.com}}.
\newblock
\newblock
\shownote{Accessed 2025}.


\bibitem[Mialon et~al\mbox{.}(2023)]%
        {Mialon2024GAIA}
\bibfield{author}{\bibinfo{person}{Gr{\'e}goire Mialon}, \bibinfo{person}{Cl{\'e}mentine Fourrier}, \bibinfo{person}{Thomas Wolf}, \bibinfo{person}{Yann LeCun}, {and} \bibinfo{person}{Thomas Scialom}.} \bibinfo{year}{2023}\natexlab{}.
\newblock \showarticletitle{{GAIA: A Benchmark for General AI Assistants}}. In \bibinfo{booktitle}{\emph{Proceedings of the Twelfth International Conference on Learning Representations}}. \bibinfo{publisher}{OpenReview.net}, \bibinfo{address}{Online}, \bibinfo{pages}{1--20}.
\newblock
\href{https://doi.org/10.48550/arXiv.2311.12983}{doi:\nolinkurl{10.48550/arXiv.2311.12983}}


\bibitem[Nie et~al\mbox{.}(2024)]%
        {nie2024hybrid}
\bibfield{author}{\bibinfo{person}{Guangtao Nie}, \bibinfo{person}{Rong Zhi}, \bibinfo{person}{Xiaofan Yan}, \bibinfo{person}{Yufan Du}, \bibinfo{person}{Xiangyang Zhang}, \bibinfo{person}{Jianwei Chen}, \bibinfo{person}{Mi Zhou}, \bibinfo{person}{Hongshen Chen}, \bibinfo{person}{Tianhao Li}, \bibinfo{person}{Ziguang Cheng}, \bibinfo{person}{Sulong Xu}, {and} \bibinfo{person}{Jinghe Hu}.} \bibinfo{year}{2024}\natexlab{}.
\newblock \showarticletitle{{A Hybrid Multi-Agent Conversational Recommender System with LLM and Search Engine in E-commerce}}. In \bibinfo{booktitle}{\emph{Proceedings of the 18th ACM Conference on Recommender Systems}}. \bibinfo{publisher}{ACM}, \bibinfo{address}{New York, NY, USA}, \bibinfo{pages}{745--747}.
\newblock
\href{https://doi.org/10.1145/3640457.3688061}{doi:\nolinkurl{10.1145/3640457.3688061}}


\bibitem[Ning et~al\mbox{.}(2024)]%
        {ning2023skeleton}
\bibfield{author}{\bibinfo{person}{Xuefei Ning}, \bibinfo{person}{Zinan Lin}, \bibinfo{person}{Zixuan Zhou}, \bibinfo{person}{Zifu Wang}, \bibinfo{person}{Huazhong Yang}, {and} \bibinfo{person}{Yu Wang}.} \bibinfo{year}{2024}\natexlab{}.
\newblock \showarticletitle{{Skeleton-of-thought: Prompting LLMs for Efficient Parallel Generation}}. In \bibinfo{booktitle}{\emph{Proceedings of the Twelfth International Conference on Learning Representations}}. \bibinfo{publisher}{OpenReview.net}, \bibinfo{address}{Online}, \bibinfo{pages}{1--20}.
\newblock
\href{https://doi.org/10.48550/arXiv.2307.15337}{doi:\nolinkurl{10.48550/arXiv.2307.15337}}


\bibitem[Park et~al\mbox{.}(2023)]%
        {park2023generative}
\bibfield{author}{\bibinfo{person}{Joon~Sung Park}, \bibinfo{person}{Joseph O'Brien}, \bibinfo{person}{Carrie~Jun Cai}, \bibinfo{person}{Meredith~Ringel Morris}, \bibinfo{person}{Percy Liang}, {and} \bibinfo{person}{Michael~S Bernstein}.} \bibinfo{year}{2023}\natexlab{}.
\newblock \showarticletitle{{Generative Agents: Interactive Simulacra of Human Behavior}}. In \bibinfo{booktitle}{\emph{Proceedings of the 2023 36th Annual ACM Symposium on User Interface Software and Technology}}. \bibinfo{publisher}{ACM}, \bibinfo{address}{New York, NY, USA}, \bibinfo{pages}{1--22}.
\newblock
\href{https://doi.org/10.1145/3586183.3606763}{doi:\nolinkurl{10.1145/3586183.3606763}}


\bibitem[Qiao et~al\mbox{.}(2024)]%
        {qiao2024autoact}
\bibfield{author}{\bibinfo{person}{Shuofei Qiao}, \bibinfo{person}{Ningyu Zhang}, \bibinfo{person}{Runnan Fang}, \bibinfo{person}{Yujie Luo}, \bibinfo{person}{Wangchunshu Zhou}, \bibinfo{person}{Yuchen~Eleanor Jiang}, \bibinfo{person}{Chengfei Lv}, {and} \bibinfo{person}{Huajun Chen}.} \bibinfo{year}{2024}\natexlab{}.
\newblock \showarticletitle{{Autoact: Automatic Agent Learning from Scratch for QA via Self-planning}}. In \bibinfo{booktitle}{\emph{Proceedings of the 62nd Annual Meeting of the Association for Computational Linguistics: Volume 1: Long Papers}}. \bibinfo{publisher}{Association for Computational Linguistics}, \bibinfo{address}{Bangkok, Thailand}, \bibinfo{pages}{3003--3021}.
\newblock
\href{https://doi.org/10.18653/v1/2024.acl-long.165}{doi:\nolinkurl{10.18653/v1/2024.acl-long.165}}


\bibitem[Rasheed et~al\mbox{.}(2024)]%
        {rasheed2024codepori}
\bibfield{author}{\bibinfo{person}{Zeeshan Rasheed}, \bibinfo{person}{Malik~Abdul Sami}, \bibinfo{person}{Kai-Kristian Kemell}, \bibinfo{person}{Muhammad Waseem}, \bibinfo{person}{Mika Saari}, \bibinfo{person}{Kari Syst{\"a}}, {and} \bibinfo{person}{Pekka Abrahamsson}.} \bibinfo{year}{2024}\natexlab{}.
\newblock \bibinfo{title}{Codepori: Large-scale System for Autonomous Software Development using Multi-agent Technology}.
\newblock
\showeprint[arxiv]{2402.01411}~[cs.SE]
\href{https://doi.org/10.48550/arXiv.2402.01411}{doi:\nolinkurl{10.48550/arXiv.2402.01411}}


\bibitem[Shaikh et~al\mbox{.}(2024)]%
        {shaikh2024rehearsal}
\bibfield{author}{\bibinfo{person}{Omar Shaikh}, \bibinfo{person}{Valentino~Emil Chai}, \bibinfo{person}{Michele Gelfand}, \bibinfo{person}{Diyi Yang}, {and} \bibinfo{person}{Michael~S Bernstein}.} \bibinfo{year}{2024}\natexlab{}.
\newblock \showarticletitle{{Rehearsal: Simulating Conflict to Teach Conflict Resolution}}. In \bibinfo{booktitle}{\emph{Proceedings of the 2024 CHI Conference on Human Factors in Computing Systems}}. \bibinfo{publisher}{ACM}, \bibinfo{address}{New York, NY, USA}, \bibinfo{pages}{1--20}.
\newblock
\href{https://doi.org/10.1145/3613904.3642159}{doi:\nolinkurl{10.1145/3613904.3642159}}


\bibitem[Shankar et~al\mbox{.}(2024)]%
        {shankar2024validates}
\bibfield{author}{\bibinfo{person}{Shreya Shankar}, \bibinfo{person}{JD Zamfirescu-Pereira}, \bibinfo{person}{Bj{\"o}rn Hartmann}, \bibinfo{person}{Aditya Parameswaran}, {and} \bibinfo{person}{Ian Arawjo}.} \bibinfo{year}{2024}\natexlab{}.
\newblock \showarticletitle{{Who Validates the Validators? Aligning LLM-assisted Evaluation of LLM Outputs with Human Preferences}}. In \bibinfo{booktitle}{\emph{Proceedings of the 2024 37th Annual ACM Symposium on User Interface Software and Technology}}. \bibinfo{publisher}{ACM}, \bibinfo{address}{New York, NY, USA}, \bibinfo{pages}{1--14}.
\newblock
\href{https://doi.org/10.1145/3654777.3676450}{doi:\nolinkurl{10.1145/3654777.3676450}}


\bibitem[Shen et~al\mbox{.}(2024)]%
        {shen2024data}
\bibfield{author}{\bibinfo{person}{Leixian Shen}, \bibinfo{person}{Haotian Li}, \bibinfo{person}{Yun Wang}, {and} \bibinfo{person}{Huamin Qu}.} \bibinfo{year}{2024}\natexlab{}.
\newblock \showarticletitle{{From Data to Story: Towards Automatic Animated Data Video Creation with LLM-based Multi-agent Systems}}. In \bibinfo{booktitle}{\emph{Proceedings of the 2024 IEEE VIS Workshop on Data Storytelling in an Era of Generative AI}}. \bibinfo{publisher}{IEEE}, \bibinfo{address}{St. Pete Beach, FL, USA}, \bibinfo{pages}{20--27}.
\newblock
\href{https://doi.org/10.1109/GEN4DS63889.2024.00008}{doi:\nolinkurl{10.1109/GEN4DS63889.2024.00008}}


\bibitem[Suzgun and Kalai(2024)]%
        {suzgun2024meta}
\bibfield{author}{\bibinfo{person}{Mirac Suzgun} {and} \bibinfo{person}{Adam~Tauman Kalai}.} \bibinfo{year}{2024}\natexlab{}.
\newblock \bibinfo{title}{Meta-prompting: Enhancing Language Models with Task-agnostic Scaffolding}.
\newblock
\showeprint[arxiv]{2401.12954}~[cs.CL]
\href{https://doi.org/10.48550/arXiv.2401.12954}{doi:\nolinkurl{10.48550/arXiv.2401.12954}}


\bibitem[Talebirad and Nadiri(2023)]%
        {talebirad2023multi}
\bibfield{author}{\bibinfo{person}{Yashar Talebirad} {and} \bibinfo{person}{Amirhossein Nadiri}.} \bibinfo{year}{2023}\natexlab{}.
\newblock \bibinfo{title}{Multi-agent Collaboration: Harnessing the Power of Intelligent LLM Agents}.
\newblock
\showeprint[arxiv]{2306.03314}~[cs.AI]
\href{https://doi.org/10.48550/arXiv.2306.03314}{doi:\nolinkurl{10.48550/arXiv.2306.03314}}


\bibitem[Tenney et~al\mbox{.}(2024)]%
        {tenney2024interactive}
\bibfield{author}{\bibinfo{person}{Ian Tenney}, \bibinfo{person}{Ryan Mullins}, \bibinfo{person}{Bin Du}, \bibinfo{person}{Shree Pandya}, \bibinfo{person}{Minsuk Kahng}, {and} \bibinfo{person}{Lucas Dixon}.} \bibinfo{year}{2024}\natexlab{}.
\newblock \bibinfo{title}{Interactive Prompt Debugging with Sequence Salience}.
\newblock
\showeprint[arxiv]{2404.07498}~[cs.CL]
\href{https://doi.org/10.48550/arXiv.2404.07498}{doi:\nolinkurl{10.48550/arXiv.2404.07498}}


\bibitem[Tran et~al\mbox{.}(2025)]%
        {tran2025multi}
\bibfield{author}{\bibinfo{person}{Khanh-Tung Tran}, \bibinfo{person}{Dung Dao}, \bibinfo{person}{Minh-Duong Nguyen}, \bibinfo{person}{Quoc-Viet Pham}, \bibinfo{person}{Barry O'Sullivan}, {and} \bibinfo{person}{Hoang~D Nguyen}.} \bibinfo{year}{2025}\natexlab{}.
\newblock \bibinfo{title}{Multi-Agent Collaboration Mechanisms: A Survey of LLMs}.
\newblock
\showeprint[arxiv]{2501.06322}~[cs.AI]
\href{https://doi.org/10.48550/arXiv.2501.06322}{doi:\nolinkurl{10.48550/arXiv.2501.06322}}


\bibitem[Triedman et~al\mbox{.}(2025)]%
        {triedman2025multiagent}
\bibfield{author}{\bibinfo{person}{Harold Triedman}, \bibinfo{person}{Rishi~Dev Jha}, {and} \bibinfo{person}{Vitaly Shmatikov}.} \bibinfo{year}{2025}\natexlab{}.
\newblock \showarticletitle{Multi-Agent Systems Execute Arbitrary Malicious Code}. In \bibinfo{booktitle}{\emph{Second Conference on Language Modeling}}. \bibinfo{publisher}{OpenReview.net}, \bibinfo{address}{Online}, \bibinfo{pages}{1--15}.
\newblock
\href{https://doi.org/10.48550/arXiv.2503.12188}{doi:\nolinkurl{10.48550/arXiv.2503.12188}}


\bibitem[Voiceflow(2025)]%
        {voiceflow}
\bibfield{author}{\bibinfo{person}{Voiceflow}.} \bibinfo{year}{2025}\natexlab{}.
\newblock \bibinfo{title}{{Voiceflow: Build AI Agents}}.
\newblock \bibinfo{howpublished}{\url{https://www.voiceflow.com/}}.
\newblock
\newblock
\shownote{Accessed: September 4, 2025}.


\bibitem[Wan et~al\mbox{.}(2024)]%
        {wan2024building}
\bibfield{author}{\bibinfo{person}{Hongyu Wan}, \bibinfo{person}{Jinda Zhang}, \bibinfo{person}{Abdulaziz~Arif Suria}, \bibinfo{person}{Bingsheng Yao}, \bibinfo{person}{Dakuo Wang}, \bibinfo{person}{Yvonne Coady}, {and} \bibinfo{person}{Mirjana Prpa}.} \bibinfo{year}{2024}\natexlab{}.
\newblock \showarticletitle{{Building LLM-based AI Agents in Social Virtual Reality}}. In \bibinfo{booktitle}{\emph{Extended Abstracts of the 2024 CHI Conference on Human Factors in Computing Systems}}. \bibinfo{publisher}{ACM}, \bibinfo{address}{New York, NY, USA}, Article \bibinfo{articleno}{65}, \bibinfo{numpages}{7}~pages.
\newblock
\href{https://doi.org/10.1145/3613905.3651026}{doi:\nolinkurl{10.1145/3613905.3651026}}


\bibitem[Wang et~al\mbox{.}(2024)]%
        {wang2024survey}
\bibfield{author}{\bibinfo{person}{Lei Wang}, \bibinfo{person}{Chen Ma}, \bibinfo{person}{Xueyang Feng}, \bibinfo{person}{Zeyu Zhang}, \bibinfo{person}{Hao Yang}, \bibinfo{person}{Jingsen Zhang}, \bibinfo{person}{Zhiyuan Chen}, \bibinfo{person}{Jiakai Tang}, \bibinfo{person}{Xu Chen}, \bibinfo{person}{Yankai Lin}, {et~al\mbox{.}}} \bibinfo{year}{2024}\natexlab{}.
\newblock \showarticletitle{{A Survey on Large Language Model based Autonomous Agents}}.
\newblock \bibinfo{journal}{\emph{Frontiers of Computer Science}} \bibinfo{volume}{18}, \bibinfo{number}{6} (\bibinfo{year}{2024}), \bibinfo{pages}{186345}.
\newblock
\href{https://doi.org/10.1007/s11704-024-40231-1}{doi:\nolinkurl{10.1007/s11704-024-40231-1}}


\bibitem[Wang et~al\mbox{.}(2025)]%
        {wang2025megaagent}
\bibfield{author}{\bibinfo{person}{Qian Wang}, \bibinfo{person}{Tianyu Wang}, \bibinfo{person}{Zhenheng Tang}, \bibinfo{person}{Qinbin Li}, \bibinfo{person}{Nuo Chen}, \bibinfo{person}{Jingsheng Liang}, {and} \bibinfo{person}{Bingsheng He}.} \bibinfo{year}{2025}\natexlab{}.
\newblock \showarticletitle{MegaAgent: A large-scale autonomous LLM-based multi-agent system without predefined SOPs}. In \bibinfo{booktitle}{\emph{Findings of the 2025 Conference of the Association for Computational Linguistics}}. \bibinfo{publisher}{Association for Computational Linguistics}, \bibinfo{address}{Vienna, Austria}, \bibinfo{pages}{4998--5036}.
\newblock
\href{https://doi.org/10.18653/v1/2025.findings-acl.259}{doi:\nolinkurl{10.18653/v1/2025.findings-acl.259}}


\bibitem[Wu et~al\mbox{.}(2023)]%
        {wu2023autogen}
\bibfield{author}{\bibinfo{person}{Qingyun Wu}, \bibinfo{person}{Gagan Bansal}, \bibinfo{person}{Jieyu Zhang}, \bibinfo{person}{Yiran Wu}, \bibinfo{person}{Beibin Li}, \bibinfo{person}{Erkang Zhu}, \bibinfo{person}{Li Jiang}, \bibinfo{person}{Xiaoyun Zhang}, \bibinfo{person}{Shaokun Zhang}, \bibinfo{person}{Jiale Liu}, {et~al\mbox{.}}} \bibinfo{year}{2023}\natexlab{}.
\newblock \bibinfo{title}{Autogen: Enabling Next-gen LLM Applications via Multi-agent Conversation}.
\newblock
\showeprint[arxiv]{2308.08155}~[cs.AI]
\href{https://doi.org/10.48550/arXiv.2308.08155}{doi:\nolinkurl{10.48550/arXiv.2308.08155}}


\bibitem[Wu et~al\mbox{.}(2024)]%
        {wu2024copilot}
\bibfield{author}{\bibinfo{person}{Zhiyong Wu}, \bibinfo{person}{Chengcheng Han}, \bibinfo{person}{Zichen Ding}, \bibinfo{person}{Zhenmin Weng}, \bibinfo{person}{Zhoumianze Liu}, \bibinfo{person}{Shunyu Yao}, \bibinfo{person}{Tao Yu}, {and} \bibinfo{person}{Lingpeng Kong}.} \bibinfo{year}{2024}\natexlab{}.
\newblock \bibinfo{title}{Os-copilot: Towards Generalist Computer Agents with Self-improvement}.
\newblock
\showeprint[arxiv]{2402.07456}~[cs.AI]
\href{https://doi.org/10.48550/arXiv.2402.07456}{doi:\nolinkurl{10.48550/arXiv.2402.07456}}


\bibitem[Yang et~al\mbox{.}(2025)]%
        {yang2025agentnet}
\bibfield{author}{\bibinfo{person}{Yingxuan Yang}, \bibinfo{person}{Huacan Chai}, \bibinfo{person}{Shuai Shao}, \bibinfo{person}{Yuanyi Song}, \bibinfo{person}{Siyuan Qi}, \bibinfo{person}{Renting Rui}, {and} \bibinfo{person}{Weinan Zhang}.} \bibinfo{year}{2025}\natexlab{}.
\newblock \bibinfo{title}{Agentnet: Decentralized Evolutionary Coordination for LLM-based Multi-Agent Systems}.
\newblock
\showeprint[arxiv]{2504.00587}~[cs.MA]
\href{https://doi.org/10.48550/arXiv.2504.00587}{doi:\nolinkurl{10.48550/arXiv.2504.00587}}


\bibitem[Zhang et~al\mbox{.}(2024)]%
        {zhang2024see}
\bibfield{author}{\bibinfo{person}{Yu Zhang}, \bibinfo{person}{Jingwei Sun}, \bibinfo{person}{Li Feng}, \bibinfo{person}{Cen Yao}, \bibinfo{person}{Mingming Fan}, \bibinfo{person}{Liuxin Zhang}, \bibinfo{person}{Qianying Wang}, \bibinfo{person}{Xin Geng}, {and} \bibinfo{person}{Yong Rui}.} \bibinfo{year}{2024}\natexlab{}.
\newblock \showarticletitle{{See Widely, Think Wisely: Toward Designing a Generative Multi-Agent System to Burst Filter Bubbles}}. In \bibinfo{booktitle}{\emph{Proceedings of the 2024 CHI Conference on Human Factors in Computing Systems}}. \bibinfo{publisher}{ACM}, \bibinfo{address}{New York, NY, USA}, \bibinfo{pages}{1--24}.
\newblock
\href{https://doi.org/10.1145/3613904.3642545}{doi:\nolinkurl{10.1145/3613904.3642545}}


\end{thebibliography}

\clearpage

\appendix
\section{Formative Study Interview Questions}\label{formative:questions}
Here, we list the pre-defined questions for the formative semi-structured interview. During the interview, if participants' initial responses did not fully cover the intended aspects of a question, we would also provide follow-up prompts accordingly.

\noindent \textit{(1) Latest Experience with Multi-Agent System Development}

\begin{itemize}
    \item[] • Please recall and describe your latest experience of building a multi-agent system.
\end{itemize}

\noindent \textit{(2) The Process of Multi-agent System Diagnosis}

\begin{itemize}
    \item[] • Did you encounter any failures in your multi-agent systems? If so, how did you diagnose the causes of the failures?

\end{itemize}

\noindent \textit{(3) The Challenges of Multi-agent System Diagnosis}

\begin{itemize}
    \item[] • Which parts of the diagnostic process did you find difficult, time-consuming, or cognitively demanding?
    \item[] • What difficulties did you face during the diagnostic process?
    \item[] • Did you use any tools to support your diagnostic process? If so, what tools or interfaces did you use? How did these tools or interfaces benefit your diagnostic process?
    \item[] • Do you have any expectations and concerns for the existing tools or interfaces for multi-agent system diagnosis?
    
\end{itemize}

\noindent \textit{(4) Others}

\begin{itemize}
    \item[] • Is there anything else about diagnosing multi-agent systems that we have not covered?
\end{itemize}

\section{User Study Interview Questions}\label{userstudy:questions}
We listed the main interview questions used in the user study. Participants were also encouraged to tell their stories and experiences outside these structured questions.

\noindent \textit{(1) User Experience During the Process}
\begin{itemize}
    \item[] • Please respectively describe your experiences completing the multi-agent system diagnosis task using the baseline system and using \tool.
    
    \item[] • Compared with the baseline system, what do you consider the main strengths and weaknesses of our proposed system?

    \item[] • How did the baseline system and \tool each affect your confidence in the identified failures?

    \item[] • How did using the baseline system and \tool influence your perceived load while completing the diagnosis task?

\end{itemize}

\noindent \textit{(2) User Perception Towards \tool}

\begin{itemize}
    \item[] • How did you perceive the usability of \tool? How did you perceive the usability of the four main features of \tool, including the activity-level summary, action-level summary, operation-level details, and filtering function?

    \item[] • How did you perceive the usefulness of \tool? How did you perceive the usefulness of the four main features of \tool, including the activity-level summary, action-level summary, operation-level details, and filtering function?

\end{itemize}

\noindent \textit{(3) Future Use and Limitations}

\begin{itemize}
    \item[] • What limitations did you observe in our system that could be addressed in the future?

    \item[] • Is there anything else you would like to add?
\end{itemize}

\end{document}